\RequirePackage{ifpdf}
\ifpdf 
\documentclass[pdftex]{sigma}
\else
\documentclass{sigma}
\fi

\numberwithin{equation}{section}
\numberwithin{theorem}{section}
\numberwithin{proposition}{section}
\numberwithin{lemma}{section}
\numberwithin{corollary}{section}
\numberwithin{definition}{section}
\numberwithin{example}{section}
\numberwithin{remark}{section}

\newcommand{\pa}{\partial}
\newcommand{\cA}{\mathcal{A}}
\newcommand{\cB}{\mathcal{B}}
\newcommand{\cE}{\mathcal{E}}
\newcommand{\cP}{\mathcal{P}}
\newcommand{\cU}{\mathcal{U}}
\newcommand{\cV}{\mathcal{V}}

\newcommand{\bA}{\boldsymbol{A}}
\newcommand{\bB}{\boldsymbol{B}}
\newcommand{\bC}{\boldsymbol{C}}
\newcommand{\bg}{\boldsymbol{g}}
\newcommand{\bGamma}{\boldsymbol{\Gamma}}
\newcommand{\bI}{\boldsymbol{I}}
\newcommand{\bL}{\boldsymbol{L}}
\newcommand{\bM}{\boldsymbol{M}}
\newcommand{\bN}{\boldsymbol{N}}
\newcommand{\bQ}{\boldsymbol{Q}}
\newcommand{\bP}{\boldsymbol{P}}
\newcommand{\bphi}{\boldsymbol{\phi}}
\newcommand{\bR}{\boldsymbol{R}}
\newcommand{\bS}{\boldsymbol{S}}
\newcommand{\bT}{\boldsymbol{T}}
\newcommand{\bU}{\boldsymbol{U}}
\newcommand{\bV}{\boldsymbol{V}}
\newcommand{\bX}{\boldsymbol{X}}
\newcommand{\bY}{\boldsymbol{Y}}
\newcommand{\bZ}{\boldsymbol{Z}}

\newcommand{\bbA}{\mathbb{A}}
\newcommand{\bbC}{\mathbb{C}}

\newcommand{\bbR}{\mathbb{R}}

\newcommand{\tphi}{\tilde{\phi}}

\newcommand{\tbP}{\tilde{\boldsymbol{P}}}
\newcommand{\tbR}{\tilde{\boldsymbol{R}}}
\newcommand{\tbX}{\tilde{\boldsymbol{X}}}

\renewcommand{\d}{\mathrm{d}}
\newcommand{\bd}{\bar{\mathrm{d}}}
\newcommand{\D}{\mathrm{D}}

\newcommand{\imag}{\mathrm{i}}

\begin{document}

\allowdisplaybreaks

\renewcommand{\PaperNumber}{118}

\FirstPageHeading

\ShortArticleName{The Non-Autonomous Chiral Model in the Bidif\/ferential Calculus Framework}

\ArticleName{The Non-Autonomous Chiral Model \\ and the
           Ernst Equation of General Relativity \\
           in the Bidif\/ferential Calculus Framework}

\Author{Aristophanes DIMAKIS~$^\dag$, Nils KANNING~$^\ddag$ and Folkert M\"ULLER-HOISSEN~$^\S$}

\AuthorNameForHeading{A.~Dimakis, N.~Kanning and F.~M\"uller-Hoissen}

\Address{$^\dag$~Department of Financial and Management Engineering, University of the Aegean,\\
\hphantom{$^\dag$}~41, Kountourioti Str., 82100 Chios, Greece}
\EmailD{\href{mailto:dimakis@aegean.gr}{dimakis@aegean.gr}}

\Address{$^\ddag$~Institute for Mathematics and Institute for Physics, Humboldt University,\\
\hphantom{$^\ddag$}~Rudower Chaussee 25, 12489 Berlin, Germany}
\EmailD{\href{mailto:kanning@mathematik.hu-berlin.de}{kanning@mathematik.hu-berlin.de}}

\Address{$^\S$~Max-Planck-Institute for Dynamics and Self-Organization,\\
 \hphantom{$^\S$}~Bunsenstrasse 10, 37073 G\"ottingen, Germany}
\EmailD{\href{mailto:folkert.mueller-hoissen@ds.mpg.de}{folkert.mueller-hoissen@ds.mpg.de}}

\ArticleDates{Received August 31, 2011, in f\/inal form December 16, 2011; Published online December 23, 2011}

\Abstract{The non-autonomous chiral model equation for an $m \times m$ matrix function on a
two-dimensional space appears in particular in general relativity, where for $m=2$ a certain
reduction of it determines stationary, axially symmetric solutions of Einstein's vacuum equations,
and for $m=3$ solutions of the Einstein--Maxwell equations.
Using a~very simple and general result of the bidif\/ferential calculus approach to integrable
partial dif\/ferential and dif\/ference equations, we generate a large class of exact solutions
of this chiral model. The solutions are parametrized by a set of matrices, the size of
which can be arbitrarily large. The matrices are subject to
a Sylvester equation that has to be solved and generically admits a unique solution.
By imposing the aforementioned reductions on the matrix data, we recover the
Ernst potentials of multi-Kerr-NUT and multi-Demia\'{n}ski--Newman metrics.}


\Classification{37K10; 16E45}

\section{Introduction}

The bidif\/ferential calculus framework allows to elaborate solution generating methods
for a~wide class of nonlinear ``integrable'' partial dif\/ferential or dif\/ference equations (PDDEs)
to a~conside\-rable extent on a universal level, i.e.\ resolved from specif\/ic examples.
It takes advantage of the simple rules underlying the calculus of dif\/ferential forms (on
a manifold), but allows for a~generalization of the latter, which is partly inspired by
noncommutative geometry.
 For a brief account of the basic structures and some results we refer to \cite{DMH08bidiff}
(also see the references therein), but all what is needed for the present work
is provided in Section~\ref{sec:basic}.
In this framework we explore the non-autonomous chiral model equation
\begin{gather}
    \big(\rho   g_z   g^{-1}\big)_z + \big(\rho  g_\rho   g^{-1}\big)_\rho  = 0  \label{naCM_g}
\end{gather}
for an $m \times m$ matrix $g$, where $\rho>0$ and $z$ are independent real variables, and a
subscript indicates a corresponding partial derivative. It apparently f\/irst appeared,
supplemented by certain reduction conditions (see Section~\ref{sec:Ernst}),
as the central part of the stationary axially symmetric Einstein vacuum ($m=2$) and
Einstein--Maxwell ($m=3$) equations (see in particular
\cite{Kinnersley77,Maison78,Maison79,Beli+Zakh78,Beli+Sakh79,Beli+Verd01,SKMHH03}).
 For $m>3$ this equation is met in higher-dimensional gravity, with a correspondingly enlarged
number of Killing vector f\/ields (see e.g.~\cite{Lee85,Lee87II,Wei01,Harm04,Pomer06,Azuma+Koikawa06,Iguchi+Mishima06,TMY06,Yazad08,Empa+Reall08,FJRV10}).
A version of the above equation also arises as the cylindrically symmetric case of the
$(2+1)$-dimensional principal chiral model~\cite{Mikh+Yare82} and as a special case of the
stationary Landau--Lifshitz equation for an isotropic two-dimensional ferromagnet \cite{Guts+Lipo95}.

The f\/irst construction of ``multi-soliton'' solutions of (\ref{naCM_g}) has been carried out
by Belinski and Zakharov \cite{Beli+Zakh78,Beli+Sakh79} (also see \cite{Beli+Verd01})
using the ``dressing method''\footnote{Further constructions of multi-soliton solutions,
in the context of general relativity, were
presented in particular by Alekseev~\cite{Alek80,Alek88}, Neugebauer~\cite{Neug80},
Kramer and Neugebauer~\cite{Neug+Kramer83,Kramer+Neug91},
Korotkin~\cite{Koro89} (limiting cases of f\/inite-gap solutions), Manko et al.~\cite{MMR95,RMM95}, and Masuda et al.~\cite{MSF98} (using Hirota's method).
Also see the references in \cite[Section~34]{SKMHH03} for other solution generating techniques.}.
Here~(\ref{naCM_g}) is expressed as the integrability condition of a linear system, which
depends on a (spectral) parameter and involves derivatives with respect to the latter.
Another approach is based on a linear system that depends on a variable spectral parameter,
i.e.\ a parameter that depends on the variables $\rho$ and $z$ \cite{Maison78}.
In Appendix~\ref{app:CM_linsys} we show that both linear systems arise from a universal
linear system (see Section~\ref{sec:basic}) in the bidif\/ferential calculus framework
(also see~\cite{BZM87} for a relation between the two linear systems).

\looseness=1
In the present work, we concentrate on a surprisingly simple general solution generating
result in the bidif\/ferential calculus framework, which has already been successfully
applied in various other cases of integrable (soliton) equations \cite{DMH08bidiff,DMH10NLS,DMH10AKNS,DKMH11}
to generate multi-soliton families.
In order to make it applicable to the non-autonomous chiral model, a slight generalization
is required, however (see Section~\ref{sec:sol_gen} and Appendix~\ref{app:gt}).
Section~\ref{sec:naCM} then elaborates it for the $m \times m$ non-autonomous chiral model.
We obtain solutions parametrized by four matrices. Two of them arise as solutions of
an $n \times n$ matrix version of the quadratic equation for pole trajectories that
f\/irst appeared in the solution generating method of Belinski and Zakharov
 \cite{Beli+Zakh78,Beli+Sakh79,Beli+Verd01}.
It then remains to solve a Sylvester equation, where two more matrices enter,
which are constant of size $m \times n$, respectively $n \times m$.
Since $n$ can be arbitrarily large, we obtain an inf\/inite family of solutions. The Sylvester
equation is easily solved if the f\/irst two matrices are chosen diagonal, and in this case
one recovers ``multi-soliton'' solutions. Additional solutions are obtained if the two
$n \times n$ matrices are \emph{non}-diagonal. In this case it is more dif\/f\/icult to solve the
Sylvester equation, though a not very restrictive spectrum condition ensures the existence
of a unique solution.
Except for an example in Section~\ref{sec:Ernst}, we will not elaborate this case further in
this work.

Section~\ref{sec:Ernst} addresses reductions, in particular to the Ernst equation of
general relativity. It turns out that the ``multi-soliton'' solutions of the stationary,
axially symmetric Einstein vacuum and Einstein--Maxwell equations are indeed in the generated
class of solutions of the non-autonomous chiral model. We thus obtain a new representation
of these solutions.
It has the property that the superposition of two (or more) ``solitons''
(e.g.\ black holes) simply corresponds to block-diagonal composition of the matrix data
parametrizing the constituents. This puts a~new perspective on an old result about
one of the most important integrable equations in physics.

We would like to stress that the solutions of the non-autonomous chiral model
and the Ernst equation(s), (re)derived in this work, originate from a \emph{universal} result
that also generates multi-soliton solutions of various other integrable equations
in a \emph{non-iterative} way.
The crucial step is to f\/ind a ``bidif\/ferential calculus formulation'' of the
respective equation. This may be regarded as a generalization of the problem of
formulating the equation as a reduction of the selfdual Yang--Mills equation.
Indeed, in the case under consideration, it is of great help that an embedding
of the non-autonomous chiral model in the ($m \times m$) selfdual Yang--Mills
equation is known \cite{Witten79,FHP80,Bais+Sasa82,Ward83,Maso+Wood96,Klein+Richt05}, and a
bidif\/ferential calculus formulation is then obtained from that of the selfdual
Yang--Mills equation \cite{DMH08bidiff}, see Section~\ref{sec:naCM}.
Once this is at hand, the remaining computations are rather straightforward.

Section~\ref{sec:conclusions} contains some concluding remarks.

\section{Preliminaries}
\label{sec:basic}
{\bf Basic def\/initions.}
A \emph{graded algebra} is an associative algebra $\Omega$ over $\bbC$ with a direct
sum decomposition $\Omega = \bigoplus_{r \geq 0} \Omega^r$
into a subalgebra $\cA := \Omega^0$ and $\cA$-bimodules $\Omega^r$, such that
$\Omega^r \, \Omega^s \subseteq \Omega^{r+s}$.
A \emph{bidifferential calculus} (or \emph{bidifferential graded algebra}) is a unital graded algebra
$\Omega$ equipped with two ($\bbC$-linear) graded derivations
$\d, \bar{\d} : \Omega \rightarrow \Omega$ of degree one (hence $\d \Omega^r \subseteq \Omega^{r+1}$,
$\bar{\d} \Omega^r \subseteq \Omega^{r+1}$), with the properties
\begin{gather}
    \d_\kappa^2 = 0    \qquad\forall \, \kappa \in \bbC   , \qquad
    \mbox{where} \quad \d_\kappa := \bd - \kappa   \d   ,
            \label{bidiff_conds}
\end{gather}
and the graded Leibniz rule
\[
    \d_\kappa(\chi  \chi') = (\d_\kappa \chi)   \chi' + (-1)^r   \chi   \d_\kappa \chi'    ,
\]
for all $\chi \in \Omega^r$ and $\chi' \in \Omega$. This means that $\d$ and $\bd$ both satisfy
the graded Leibniz rule. In Section~\ref{sec:sol_gen} we consider a more narrow class of
graded algebras. A bidif\/ferential calculus within this class is then specif\/ied in Section~\ref{sec:naCM}.

{\bf Dressing a bidif\/ferential calculus.}
Let $(\Omega, \d, \bd)$ be a bidif\/ferential calculus. Repla\-cing~$\d_\kappa$ in (\ref{bidiff_conds}) by
\begin{gather*}
     \D_\kappa := \bd - \bbA - \kappa   \d     ,
\end{gather*}
with a 1-form $\bbA$ (i.e.\ an element of $\Omega^1$), the resulting condition $\D_\kappa^2 =0$
(for all $\kappa \in \bbC$) can be expressed as
\begin{gather}
    \d \bbA = 0  \qquad \mbox{and} \qquad   \bd \bbA - \bbA   \bbA = 0   .
                          \label{gauged-bdga-cond}
\end{gather}
If these equations are equivalent to a PDDE or a system of PDDEs for a set of functions,
we say we have a \emph{bidifferential calculus formulation} for it. This requires that
$\bbA$ depends on these functions and the derivations $\d$, $\bd$
involve dif\/ferential or dif\/ference operators. There are several ways to reduce the two equations
(\ref{gauged-bdga-cond}) to a single one. Here we only consider two of them.

\textbf{1.} We can solve the f\/irst of (\ref{gauged-bdga-cond}) by setting
\begin{gather*}
     \bbA = \d \phi  .     
\end{gather*}
This converts the second of (\ref{gauged-bdga-cond}) into
\begin{gather}
     \bd   \d   \phi = \d\phi  \, \d\phi   .    \label{phi_eq}
\end{gather}
This equation is obviously invariant under $\phi \mapsto \alpha \phi \alpha^{-1} + \beta$ with
an invertible $\alpha \in \cA$ satisfying $\d \alpha = \bd \alpha =0$, and $\beta \in \cA$
satisfying $\d \beta =0$.

\textbf{2.} Alternatively, the second of equations (\ref{gauged-bdga-cond}) can be solved by setting
\begin{gather*}
      \bbA = (\bd g)  g^{-1}   ,   
\end{gather*}
and the f\/irst equation then reads
\begin{gather}
        \d   \big( (\bd g)   g^{-1} \big) =0   .   \label{g_eq}
\end{gather}
This equation has the (independent left and right handed, i.e.\ chiral) symmetry
\begin{gather}
     g \mapsto \alpha   g   \beta   ,    \label{g_eq_sym}
\end{gather}
where $\alpha \in \cA$ is $\d$-constant\footnote{Although not evident, we need not
require $\bd \alpha=0$ in addition.}
and $\beta \in \cA$ is $\bd$-constant, and both have to be invertible.
Since
\begin{gather}
    \bd   \big[ \big(\d g^{-1}\big)   g \big]
  = - \bd \big( g^{-1} \d g \big) = g^{-1}   \big[ (\d \bd g)   g^{-1} - (\bd g)   \d g^{-1} \big]   g
  = g^{-1}   \d \big[ (\bd g)   g^{-1} \big]   g   ,    \label{d,bd_g->g^-1}
\end{gather}
$g$ solves (\ref{g_eq}) if\/f $g^{-1}$ solves (\ref{g_eq}) with $\d$ and $\bd$ exchanged.
In our central example, the non-autonomous chiral model, $g \mapsto g^{-1}$ becomes a symmetry.

{\bf Linear system.} The compatibility condition of the \emph{linear} equation
\begin{gather}
    \bd X = (\d X)   P + \bbA   X     \label{bidiff_linsys}
\end{gather}
is
\[
    0 = \bd^2 X = (\d X)   [ (\d P) P - \bd P ] + \big(\bd \bbA -\bbA^2\big)   X - (\d \bbA)   X   P    .
\]
If $P$ satisf\/ies
\begin{gather}
     \bd P = (\d P)   P   ,   \label{bidiff_P_eq}
\end{gather}
this reduces to
\begin{gather}
      \big(\bd \bbA - \bbA^2\big)   X = (\d \bbA)   X   P  .  \label{bidiff_linsys_integr}
\end{gather}
For the above choices of $\bbA$, this implies the respective PDDE. Hence (\ref{bidiff_linsys})
is the source of a~corresponding Lax pair, also see Appendix~\ref{app:CM_linsys}.

As a consequence of (\ref{bidiff_P_eq}), $P$ solves (\ref{phi_eq}) and,
if $P$ is invertible, also (\ref{g_eq}).

{\bf Miura transformation.}
If a pair $(\phi,g)$ solves the \emph{Miura transformation} equation
\begin{gather}
     (\bd g)   g^{-1} = \d \phi  \label{Miura}
\end{gather}
(cf.~\cite{DMH08bidiff}), it follows (as an integrability condition) that $\phi$ solves
(\ref{phi_eq}) and $g$ solves (\ref{g_eq}). We note that (\ref{Miura}) is just
the linear equation (\ref{bidiff_linsys}) if we identify $\bbA=\d \phi$, $X=g$ and set $P=0$.
If we have chosen a
bidif\/ferential calculus and a reduction condition such that (\ref{phi_eq})
becomes equivalent to some PDDE, this does not necessarily mean that also (\ref{g_eq})
is equivalent to some ``ordinary'' PDDE. But for the central example of this work, the non-autonomous
chiral model, such a~mismatch does not occur. In fact, in Section~\ref{sec:sol_gen}
we will actually present a solution generating method for (\ref{Miura}).

\section{A solution generating method}\label{sec:sol_gen}

Let $\bigwedge(\bbC^N)$ denote the exterior (Grassmann) algebra
of the vector space $\bbC^N$ and $\mathrm{Mat}(m,n,\cB)$ the set of $m \times n$ matrices
with entries in some unital algebra $\cB$. We choose $\cA$ as the algebra of all
f\/inite-dimensional matrices (with entries in $\cB$), where the product of two matrices is
def\/ined to be zero if the sizes of the two matrices do not match, and assume that
$\Omega = \cA \otimes \bigwedge(\bbC^N)$ is supplied with the structure of a bidif\/ferential calculus.
In the following, $I=I_m$ and $\bI=\bI_n$ denote the $m \times m$, respectively $n \times n$,
identity matrix.

\begin{proposition}
\label{prop:main}
Let $\bP,\bR,\bX \in \mathrm{Mat}(n,n,\cB)$ be invertible solutions of
\begin{gather}
   \bar{\d} \bP = (\d \bP)   \bP   , \qquad
   \bd \bR = \bR   \d \bR   ,\nonumber \\
   \bar{\d} \bX = (\d \bX)   \bP - (\d \bR)   \bX    , \qquad
   \bX   \bP - \bR   \bX = \bV  \bU   ,  \label{prop_conds}
\end{gather}
with $\d$- and $\bar{\d}$-constant $\bU \in \mathrm{Mat}(m,n,\cB)$, $\bV \in \mathrm{Mat}(n,m,\cB)$.
Then
\begin{gather}
    \phi = \bU \bX^{-1} \bV   , \qquad
       g = I + \bU (\bR   \bX)^{-1} \bV    \label{prop_phi,g}
\end{gather}
solve the Miura transformation equation \eqref{Miura}, and thus \eqref{phi_eq},
respectively \eqref{g_eq}.
\end{proposition}

\begin{proof}
Using the last three of (\ref{prop_conds}) we obtain
\begin{gather*}
   \bd (\bR \bX)^{-1}  =  - \bX^{-1} \big[ \bd \bX   \bX^{-1} + \bR^{-1} \bd \bR \big]   \bR^{-1}
    =  - \bX^{-1} (\d \bX)   \bX^{-1} (\bX \bP)   (\bR \bX)^{-1} \\
\phantom{\bd (\bR \bX)^{-1} }{} =  \big(\d \bX^{-1}\big) [ \bI + \bV   \bU   (\bR \bX)^{-1} ]   .
\end{gather*}
Multiplication by $\bU$ from the left and by $\bV$ from the right, and using $\bd I =0$, leads to
\[
    \bd g = \bU (\d \bX^{-1}) \bV   g = (\d \phi)   g   .
\]
Hence $\phi$ and $g$ solve the Miura transformation equation (\ref{Miura}).
We did not use the f\/irst of (\ref{prop_conds}), but it arises as an integrability
condition: $0 = \bd^2 \bX = (\d \bX)   [(\d \bP)   \bP - \bd \bP ]$.
\end{proof}

\begin{remark}
\label{rem:Sylv->X-eq}
The third of (\ref{prop_conds}), which has the form of the linear equation (\ref{bidiff_linsys}),
is almost a consequence of the fourth, which is a Sylvester equation.
Indeed, as a consequence of the Sylvester equation we have
\begin{gather*}
    0  =  \bd ( \bR  \bX - \bX  \bP + \bV  \bU )
       = (\bd \bR)  \bX + \bR  \bd \bX - (\bd \bX)  \bP - \bX  \bd \bP  \\
\phantom{0}{} =  \bR  [ \bd \bX + (\d \bR)  \bX ] - [ \bd \bX + \bX  \d \bP ]  \bP \\
\phantom{0}{}      =  \bR  [ \bd \bX + (\d \bR)  \bX - (\d \bX)  \bP]
          - [ \bd \bX + (\d \bR)  \bX - (\d \bX)  \bP ]  \bP
          + \d ( \bR  \bX - \bX  \bP )  \bP   ,
\end{gather*}
where the last term vanishes.
If $\bP$ and $\bR$ are suf\/f\/iciently independent, this implies that the third of
(\ref{prop_conds}) is satisf\/ied. In particular, this holds if $\cB$ is the algebra of
complex functions of some variables and if $\bP$ and $\bR$ have no eigenvalue in common.
\end{remark}

Appendix~\ref{app:gt} explains how Proposition~\ref{prop:main} arises from a theorem
that has been applied in previous work to generate soliton solutions of several integrable
PDDEs.

\section{The non-autonomous chiral model}
\label{sec:naCM}

The PDE def\/ining the non-autonomous chiral model can be obtained as a reduction of the
self-dual Yang--Mills (sdYM) equation (see e.g.~\cite{Witten79,FHP80,Bais+Sasa82,Ward83,Maso+Wood96,Klein+Richt05}).
In an analogous way, a~bidif\/ferential calculus for the non-autonomous chiral model can be
derived from a bidif\/ferential calculus
for the sdYM equation (also see~\cite{Kanning10}). In coordinates $\rho$, $z$, $\theta$,
where $\rho>0$, it is given by
\begin{gather}
    \d f = - f_z  \zeta_1 + e^\theta  \big(f_\rho - \rho^{-1} f_\theta\big)  \zeta_2  , \qquad
    \bar{\d} f = e^{-\theta}  \big(f_\rho + \rho^{-1} f_\theta\big)  \zeta_1 + f_z  \zeta_2   .
          \label{CM_bidiff}
\end{gather}
Here e.g.~$f_z$ denotes the partial derivative of a function $f$ (of the three coordinates)
with respect to~$z$, and $\zeta_1$, $\zeta_2$ is a basis of $\bigwedge^1(\mathbb{C}^2)$.
$\d$ and $\bd$ extend to matrices of functions and moreover to
$\Omega = \cA \otimes \bigwedge(\bbC^2)$ with
$\cA = \mathrm{Mat}(m,m,\bbC)$, treating $\zeta_1$, $\zeta_2$ as constants.
The coordinate $\theta$ is needed to have the properties of a bidif\/ferential calculus,
but we are f\/inally interested in equations for objects that do not depend on it.

A (matrix-valued) function is $\d$-constant ($\bd$-constant) if\/f it is $z$-independent and only depends on the
variables $\theta$, $\rho$ through the combination $\rho   e^\theta$ (respectively
$\rho   e^{-\theta}$).
It is $\d$- and $\bd$-constant if\/f it is constant, i.e.\ independent of $z$, $\theta$, $\rho$.

For an $m \times m$ matrix-valued function $g$, (\ref{g_eq}) takes the form
\[
    \big(\rho   g_z   g^{-1}\big)_z + \big(\rho   g_\rho   g^{-1}\big)_\rho
    - \big(g_\rho  g^{-1}\big)_\theta + \big(g_\theta   g^{-1}\big)_\rho
    - \rho^{-1} \big(g_\theta   g^{-1}\big)_\theta  = 0  .
\]
Restricting $g$ by setting
\begin{gather*}
     g = e^{c   \theta}   \tilde{g}   
\end{gather*}
with any constant $c$ and $\theta$-independent $\tilde{g}$, for the latter we obtain the
\emph{non-autonomous chiral model} equation\footnote{Changing the sign of the f\/irst term in
the expression for $\d f$ in~(\ref{CM_bidiff}), we obtain a minus sign between the two
terms on the left hand side of~(\ref{CM}). This hyperbolic version of the chiral model shows up,
in particular, in the reduction of the Einstein vacuum equations with two spacelike
commuting Killing vector f\/ields, describing gravitational plane waves~\cite{Beli+Verd01}.
Our further analysis can be adapted to this case.}
\begin{gather}
    \big(\rho  \tilde{g}_z   \tilde{g}^{-1}\big)_z + \big(\rho   \tilde{g}_\rho   \tilde{g}^{-1}\big)_\rho  = 0   .
                 \label{CM}
\end{gather}
In Section~\ref{subsec:CM_sol}, we derive a family of exact solutions by application of
Proposition~\ref{prop:main}. In Appendix~\ref{app:CM_linsys} we recover two familiar linear
systems (Lax pairs) for this equation.

{\bf Miura transformation.}
Evaluating (\ref{phi_eq}) with
\begin{gather*}
     \phi = e^{-\theta}   \tphi   ,  
\end{gather*}
where $\tphi$ is $\theta$-independent, we obtain
\begin{gather}
    \tphi_{zz} + \tphi_{\rho\rho} + \rho^{-1} \tphi_\rho = \big[\tphi_\rho + \rho^{-1} \tphi , \tphi_z \big]
    ,   \label{CM_Miura_dual}
\end{gather}
which is related to the non-autonomous chiral model by the Miura transformation
\begin{gather*}
    \tphi_z = - \tilde{g}_\rho   \tilde{g}^{-1} - c   \rho^{-1}   I   , \qquad
    \tilde{g}_z   \tilde{g}^{-1} = \tphi_\rho + \rho^{-1}   \tphi   .  
\end{gather*}

{\bf Symmetries.}
(\ref{CM}) is invariant under each of the following transformations, and thus, more generally,
any combination of them.
\begin{enumerate}\itemsep=0pt
\item[(1)] $\tilde{g} \mapsto \alpha   \tilde{g}   \beta$,
with any invertible constant $m \times m$ matrices $\alpha$ and $\beta$ (cf.~(\ref{g_eq_sym})).
\item[(2)] $\tilde{g} \mapsto \rho^c   \tilde{g}$ with any constant $c$.
\item[(3)] $\tilde{g} \mapsto \tilde{g}^{-1}$ (also see (\ref{d,bd_g->g^-1})).
\item[(4)] $\tilde{g} \mapsto \tilde{g}^\dagger$, where ${}^\dagger$ indicates Hermitian conjugation.
\end{enumerate}
We note that $\tilde{g} \mapsto (\tilde{g}^\dagger)^{-1}$ is a fairly obvious symmetry. With its help,
(4) follows immediately from~(3).

\subsection{A family of exact solutions}
\label{subsec:CM_sol}
Let us f\/irst consider the equation $\bd \bP = (\d \bP)   \bP$, which is the f\/irst
of (\ref{prop_conds}). Using the above bidif\/ferential
calculus, it takes the form
\[
     \bP_z   \bP = - e^{-\theta} \big(\bP_\rho + \rho^{-1} \bP_\theta\big)   , \qquad
     \bP_z = e^{\theta} \big(\bP_\rho - \rho^{-1} \bP_\theta\big)   \bP   .
\]
Writing
\[
      \bP = e^{-\theta} \tbP    ,
\]
and assuming that $\tbP$ does not depend on $\theta$, this translates to
\begin{gather}
     \tbP_\rho - \rho^{-1} \tbP = -\tbP_z   \tbP     , \qquad
     \tbP_z = \big(\tbP_\rho + \rho^{-1} \tbP\big)   \tbP   . \label{tbP_eqs}
\end{gather}
The proof of the following result is provided in Appendix~\ref{app:proofs}.

\begin{lemma}
\label{lemma:tbP-id}
The following holds.
\begin{enumerate}\itemsep=0pt
\item[$(1)$] If $\tbP$ and $\bI + \tbP^2$ are invertible, the system \eqref{tbP_eqs} implies
\begin{gather}
    \tbP^2 - 2 \rho^{-1}   (z   \bI + \bB)   \tbP - \bI = 0  ,  \label{tbP_id}
\end{gather}
with a constant matrix $\bB$.
\item[$(2)$] Let $\bI + \tbP^2$ be invertible and $\tbP_\rho$, $\tbP_z$ commute with~$\tbP$.
If~$\tbP$ satisfies~\eqref{tbP_id}, then $\tbP$ solves~\eqref{tbP_eqs}.
\end{enumerate}
\end{lemma}

\begin{remark}
If $\tbP$ is \emph{diagonal}, then (\ref{tbP_id}) becomes the set of quadratic equations~(2.11) in~\cite{Beli+Sakh79} (or~(1.67) in~\cite{Beli+Verd01}), which determine the
``pole trajectories'' in the framework of Belinski and Zakharov. In our approach,
there are more solutions since $\tbP$ need not be diagonal.
\end{remark}

\begin{remark}
\label{rem:simple_spec}
The conditions $[\tbP_\rho , \tbP] = [\tbP_z , \tbP] = 0$ in part (2) of the lemma are
satisf\/ied in particular if the spectrum $\mathrm{spec}(\bB)$ is simple, i.e.\ if
the eigenvalues of $\bB$ are all distinct, since then
the solutions of (\ref{tbP_id}) are functions of $\rho$, $z$ and the matrix $\bB$ (and thus
$\tbP_\rho$ and $\tbP_z$ commute with $\tbP$) \cite{Higham87}. But this would be
unnecessarily restrictive, see Section~\ref{subsec:CM_Jordan}.
\end{remark}

\begin{remark}
Under the assumption that $\bI + \tbP^2$ is invertible, (\ref{tbP_eqs})
implies $[\tbP_\rho , \tbP] {=} [\tbP_z , \tbP] {=} 0$, also see~(\ref{tbP_eqs_decoupled}).
 For the bidif\/ferential calculus under consideration, $\bd \bP = (\d \bP)   \bP$ is
therefore equivalent to $\bd \bP = \bP   \d \bP$.
The latter is one of our equations for $\bR$ in Proposition~\ref{prop:main}.
Setting
\[
      \bR = e^{-\theta} \tbR    ,
\]
with $\tbR$ $\theta$-independent, invertible $\tbP$ and $\tbR$ both have to
solve~(\ref{tbP_id}).
\end{remark}

The third of (\ref{prop_conds}) becomes
\[
    \bX_\rho + \rho^{-1} \bX_\theta = - \bX_z   \tbP + \tbR_z   \bX   , \qquad
    \bX_z = \big(\bX_\rho - \rho^{-1} \bX_\theta \big)   \tbP - \big(\tbR_\rho + \rho^{-1} \tbR\big)   \bX   .
\]
Assuming that $\bU$ and $\bV$ are $\theta$-independent, and recalling the $\theta$-dependence
of~$\phi$, the formula for $\phi$ in (\ref{prop_phi,g}) requires $\bX = e^\theta \tbX$ with
$\theta$-independent $\tbX$.
Hence
\begin{gather}
    \tbX_\rho + \rho^{-1} \tbX = - \tbX_z   \tbP + \tbR_z  \tbX   , \qquad
    \tbX_z = \big(\tbX_\rho - \rho^{-1} \tbX \big)  \tbP - (\tbR_\rho + \rho^{-1} \tbR)   \tbX   .
        \label{CM_cor_Xeqs}
\end{gather}
The last of (\ref{prop_conds}) becomes the $\theta$-independent Sylvester equation
\begin{gather}
    \tbX   \tbP - \tbR   \tbX = \bV \bU   .  \label{CM_Sylv}
\end{gather}
Now Proposition~\ref{prop:main} implies the following.

\begin{proposition}
\label{prop:CM_sol}
Let $n \times n$ matrices $\tbP$ and $\tbR$ be solutions of \eqref{tbP_id} $($with a matrix~$\bB$,
respectively $\bB')$, with the properties that they commute with their derivatives w.r.t.~$\rho$
and $z$, and that $\bI + \tbP^2$ and $\bI + \tbR^2$ are invertible.
Furthermore, let $\mathrm{spec}(\tbP) \cap \mathrm{spec}(\tbR) = \varnothing$
and $\tbX$ an invertible solution of the Sylvester equation~\eqref{CM_Sylv}
with constant $m \times n$, respectively $n \times m$, matrices $\bU$ and~$\bV$. Then
\begin{gather}
    \tilde{g} = \big(I + \bU (\tbR \tbX)^{-1} \bV\big)   g_0   ,   \label{CM_sol}
\end{gather}
with any constant invertible $m \times m$ matrix\footnote{Here $g_0$ represents the
freedom of chiral transformations.}
$g_0$, solves the non-autonomous chiral model equation~\eqref{CM}.
\end{proposition}

\begin{proof} As a consequence of the spectrum condition, a solution $\tbX$ of the Sylvester
equation~(\ref{CM_Sylv}) exists and is unique. The further assumptions for $\tbP$ and $\tbR$
are those of Lemma~\ref{lemma:tbP-id}, part~(2). Furthermore, (\ref{CM_cor_Xeqs})
is a consequence of (\ref{CM_Sylv}) if the spectrum condition holds (also see
Remark~\ref{rem:Sylv->X-eq}).
Now our assertion follows from Proposition~\ref{prop:main} and the preceding
calculations.
\end{proof}

\begin{remark}
\label{rem:CM_sol_det}
The determinant of (\ref{CM_sol}) is obtained via Sylvester's theorem,
\begin{gather*}
     \det(\tilde{g})
  =  \det\big( I + \bU (\tbR \tbX)^{-1} \bV \big)   \det(g_0)
  = \det\big( \bI + \bV \bU (\tbR \tbX)^{-1} \big)   \det(g_0) \\
\phantom{\det(\tilde{g})}{}  =  \det( \tbR \tbX + \bV \bU )   \det(\tbR \tbX)^{-1}  \det(g_0)
  = \det( \tbX \tbP )  \det(\tbR \tbX)^{-1}   \det(g_0) \\
\phantom{\det(\tilde{g})}{} = \frac{\det(\tbP)}{\det(\tbR)}   \det(g_0)   ,
\end{gather*}
where we used the Sylvester equation (\ref{CM_Sylv}) and assumed that it has an invertible
solution.
\end{remark}

\begin{remark}
\label{rem:CM_sol_transf}
As an obvious consequence of (\ref{CM_Sylv}), $\bU$ and $\bV$ enter $\tilde{g}$ given by
(\ref{CM_sol}) only modulo an arbitrary scalar factor dif\/ferent from zero. We also note
that a transformation
\begin{gather*}
    \tbP \mapsto \bT_1^{-1} \tbP   \bT_1   ,\! \qquad
    \tbR \mapsto \bT_2^{-1} \tbR   \bT_2   ,\! \qquad
    \bU \mapsto \bU \bT_1   , \!\qquad
    \bV \mapsto \bT_2^{-1} \bV    , \!\qquad
    \tbX \mapsto \bT_2^{-1} \tbX \bT_1   ,
\end{gather*}
with constant invertible $n \times n$ matrices $\bT_1$, $\bT_2$,
leaves (\ref{tbP_id}), (\ref{CM_cor_Xeqs}), (\ref{CM_Sylv}) and (\ref{CM_sol}) invariant.
As a consequence, without restriction of generality, we can assume that the matrix $\bB$ in
(\ref{tbP_id}), and the corresponding matrix related to $\tbR$, both have Jordan normal form.
\end{remark}

\begin{example}
\label{ex:CM_diag}
Let $\tbP$ and $\tbR$ be diagonal, i.e.
\[
     \tbP = \left( p_i  \delta_{ij} \right)   , \qquad
     \tbR = \left( r_i  \delta_{ij} \right)   .
\]
If they have no eigenvalue in common, then (\ref{CM_Sylv}) has a unique solution
given by the Cauchy-like matrix
\[
    \tbX_{ij} = \frac{(\bV \bU)_{ij}}{p_j - r_i}   .
\]
It remains to solve (\ref{tbP_id}) (choosing $\bB$ diagonal), which yields
\begin{gather}
   p_i = \rho^{-1}   \Big( z + b_i + j_i   \sqrt{(z+b_i)^2 + \rho^2} \Big)   , \qquad
   r_i = \rho^{-1}   \Big( z + b_i' + j_i'  \sqrt{(z+b_i')^2 + \rho^2} \Big)   ,
         \label{p_i,r_i}
\end{gather}
with constants $b_i$, $b_i'$ and $j_i, j_i' \in \{\pm 1\}$. Since we assume that
$\{p_i\} \cap \{r_i\} = \varnothing$,
the assumptions of Proposition~\ref{prop:CM_sol} are satisf\/ied.
It follows that, with the above data, (\ref{CM_sol}) solves the non-autonomous chiral
model equation.
\end{example}

The case where $\tbP$ or $\tbR$ is non-diagonal is exploited in the next subsection.
But Example~\ref{ex:CM_diag} will be suf\/f\/icient to understand most of Section~\ref{sec:Ernst}.

\subsection{More about the family of solutions}\label{subsec:CM_Jordan}

Introducing matrices $\bA$ and $\bL$ via
\[
     \bA = ( z   \bI + \bB )^2 + \rho^2 \bI    , \qquad
     \tbP = \rho^{-1}   ( \bL + z   \bI + \bB )    ,
\]
(\ref{tbP_id}) translates into
\begin{gather*}
       \bL^2 = \bA  . 
\end{gather*}
According to Remark~\ref{rem:CM_sol_transf}, we can take $\bB$ in Jordan normal form,
\[
     \bB = \mbox{block-diag}(\bB_{n_1},\ldots,\bB_{n_s})   .
\]
Let us f\/irst consider the case where $\bB$ is a single $r \times r$ Jordan block,
\[
       \bB_r = b \, \bI_r + \bN_r    , \qquad
       \bN_r = \begin{pmatrix} 0 & 1 & 0 &   \cdots & 0 \\
                                          0 & 0 & 1 & \ddots         & \vdots \\
                                     \vdots &   & \ddots & \ddots  & \vdots \\
                                   \vdots &   &   &         \ddots & 1  \\
                                        0 & \cdots & \cdots  & \cdots & 0
                    \end{pmatrix} .
\]
Then we have
\[
    \bA = \mathfrak{r}^2   (\bI_r + \bM_r)   ,
\]
where
\[
    \bM_r = \mathfrak{r}^{-2} \big[ 2(z+b)   \bN_r + \bN_r^2 \big]   , \qquad
    \mathfrak{r} = \pm \sqrt{(z+b)^2 + \rho^2}   ,
\]
and thus
\[
    \bL = \mathfrak{r}   (\bI_r + \bM_r)^{1/2}
        = \mathfrak{r}   \sum_{k=0}^{r-1} {1/2 \choose k}   \bM_r^k  ,
\]
by use of the generalized binomial expansion formula, noting that $\bM_r^r =0$ as a consequence
of $\bN_r^r=0$. Hence we obtain
the following solution of (\ref{tbP_id}),
\begin{gather}
    \tbP_r = \rho^{-1}   \left( z   \bI_r + \bB_r + \mathfrak{r}   \sum_{k=0}^{r-1} {1/2 \choose k}   \bM_r^k \right)
              ,   \label{P-block_from_B-Jordan}
\end{gather}
which is an upper triangular Toeplitz matrix. In particular, we have
\begin{gather*}
   \tbP_1  =  \rho^{-1} \, [ z + b + \mathfrak{r} ]   , \\
   \tbP_2  =  \rho^{-1} [ z + b + \mathfrak{r} ]
        \begin{pmatrix} 1 & \mathfrak{r}^{-1} \\
           0 & 1 \end{pmatrix} , \\
   \tbP_3  =  \rho^{-1} [ z + b + \mathfrak{r} ]
        \begin{pmatrix} 1 & \mathfrak{r}^{-1} &
           \frac{1}{2} \rho^2 \mathfrak{r}^{-3} (z+b + \mathfrak{r})^{-1} \\
           0 & 1 & \mathfrak{r}^{-1} \\
           0 & 0 & 1 \end{pmatrix}   ,
\\
    \tbP_4  =  \rho^{-1} [ z + b + \mathfrak{r} ]
        \begin{pmatrix} 1 & \mathfrak{r}^{-1} &
           \frac{1}{2} \rho^2 \mathfrak{r}^{-3} (z+b + \mathfrak{r})^{-1}
             & -\frac{1}{2} (z+b) \, \rho^2 \mathfrak{r}^{-5} (z+b + \mathfrak{r})^{-1} \\
           0 & 1 & \mathfrak{r}^{-1} & \frac{1}{2} \rho^2 \mathfrak{r}^{-3}
                   (z+b + \mathfrak{r})^{-1} \\
           0 & 0 & 1 & \mathfrak{r}^{-1} \\
           0 & 0 & 0 & 1  \end{pmatrix}   .
\end{gather*}
These matrices are obviously nested and, from one to the next, only the entry in the right upper
corner is new.

 For the above Jordan normal form of $\bB$, solutions of (\ref{tbP_id}) are now given by\footnote{For $\tbP$ with \emph{simple} spectrum, every solution of
(\ref{tbP_id}) has this form. Otherwise, i.e.\ when there are two
Jordan blocks with the same eigenvalue, (\ref{tbP_id}) has additional solutions, see~\cite{Higham87}.
But they are further constrained by (\ref{tbP_eqs}). We will not consider such solutions in
this work.}
\begin{gather*}
     \tbP = \mbox{block-diag}(\tbP_{n_1},\ldots,\tbP_{n_s})   ,   
\end{gather*}
where the blocks typically involve dif\/ferent constants replacing $b$, i.e.\ dif\/ferent eigenvalues of~$\bB$.
Since $\tbP_\rho$ and $\tbP_z$ obviously commute with $\tbP$, and since $\bI+\tbP^2$ is
invertible\footnote{Note that
$I_r + \tbP_r^2 = (1 + \rho^{-2} (z+b+\mathfrak{r})^2)  I + \sum\limits_{k=1}^{r-1} g_k
\bN_r^k$ with functions $g_k$.} for $\rho >0$, Lemma~\ref{lemma:tbP-id}, part~(2),
ensures that $\tbP$ solves~(\ref{tbP_eqs}).
If $\tbP$ has the above form, and~$\tbR$ a similar form, and if~$\tbP$ and~$\tbR$ have disjoint
spectra, it remains to solve the Sylvester equation\footnote{Under the stated conditions the
Sylvester equation possesses a unique solution and a vast literature exists to express it.}
(\ref{CM_Sylv}) in order that~(\ref{CM_sol}) yields solutions of the non-autonomous chiral model equation.
This leads to a~plethora of exact solutions. We postpone an example to Section~\ref{sec:Ernst},
where additional conditions considerably reduce the freedom we have here, see
Example~\ref{ex:Ernst_P-Jordan_R-diag}.

\section{Reductions of the non-autonomous chiral model\\ to Ernst equations}
\label{sec:Ernst}

According to Section~\ref{sec:naCM}, a particular involutive symmetry of the non-autonomous
chiral mo\-del~(\ref{CM}) is given by $\tilde{g} \mapsto \gamma \, (\tilde{g}^\dagger)^{-1} \gamma$,
where $\gamma$ is a constant matrix with
\[
      \gamma^\dagger = \gamma   , \qquad   \gamma^2 = I   .
\]
(\ref{CM}) therefore admits the generalized unitarity reduction
$\tilde{g}^\dagger   \gamma   \tilde{g} = \gamma$,
which means that $\tilde{g}$ belongs to the unitary group $U(m;\gamma)$.\footnote{If
$\gamma$ has $p$ positive and $q$ negative eigenvalues, this is commonly denoted $U(p,q)$.}
Another reduction, associated with an involutive symmetry, is
$\tilde{g}^\dagger = \tilde{g}$.
Imposing both reductions simultaneously, amounts to setting
\begin{gather}
      \tilde{g}^\dagger = \tilde{g}   , \qquad (\gamma \tilde{g})^2 = I   .
            \label{red_cond1}
\end{gather}
Writing
\[
      \tilde{g} = \gamma   (I - 2   \cP)   ,
\]
translates these conditions into
\[
      \gamma \cP^\dagger \gamma = \cP   , \qquad \cP^2 = \cP   .
\]
In particular, $\cP$ is a projector. If we require in addition that $\mathrm{rank}(\cP)=1$,
which for a projector is equivalent to $\mathrm{tr}(\cP)=1$ \cite[Fact~5.8.1]{Bern09},
the following parametrization of $\tilde{g}$ can be achieved (also see e.g.~\cite{Zakh+Mikh78rel,Eich+Forg80,Mazur82,Guerses84}),
\begin{gather}
     \tilde{g} = \gamma - 2 \frac{v  v^\dagger}{v^\dagger \gamma  v} , \label{g_param}
\end{gather}
where $v$ is an $m$-component vector with $v^\dagger \gamma  v \neq 0$. This parametrization
is invariant under $v \mapsto c  v$ with a nowhere vanishing function $c$, so that the
f\/irst component of $v$ can be set to $1$ in the generic case where it is dif\/ferent from zero.
If $\gamma$ has signature $m-1$, (\ref{g_param}) is a parametrization of the symmetric space
$SU(m-1,1)/S(U(m-1) \times U(1))$ \cite{Zakh+Mikh78rel,Eich+Forg80,Mazur82}.
The condition $\mathrm{tr}(\cP)=1$ corresponds to
\begin{gather}
       \mathrm{tr}(\gamma \tilde{g}) = m-2   .    \label{red_cond2}
\end{gather}
We also note that $\det(\tilde{g}) = - \det(\gamma)$.
The following result, which we prove in Appendix~\ref{app:proofs}, shows how the reduction
conditions~(\ref{red_cond1}) and (\ref{red_cond2}) can be implemented on
the family of solutions of the non-autonomous chiral model obtained via
Proposition~\ref{prop:CM_sol}.

\begin{proposition}
\label{prop:CMred_quadr_constr}
Let $\tbX$ solve the Sylvester equation \eqref{CM_Sylv}, where $\tbP$, $\tbR$, $\bU$, $\bV$,
satisfy\footnote{These conditions are motivated by
the structure of $\tbP$, $\tbR$, $\bU$, $\bV$ found in Example~\ref{ex:Kerr-NUT}.}
\begin{gather}
     (\bGamma   \tbP)^2 = -\bI   , \qquad
     (\bGamma   \tbR)^2 = -\bI   , \qquad
     g_0 \gamma   \bU = \bU   \bGamma   , \qquad
     \bGamma   \bV = \bV   g_0 \gamma     ,   \label{Gg_constraints}
\end{gather}
with an $n \times n$ matrix $\bGamma$ and a constant $m \times m$ matrix $g_0$ satisfying
\begin{gather}
        \bGamma^2=\bI   , \qquad (g_0 \gamma)^2 = I   .  \label{Gg0_conds}
\end{gather}
Furthermore, let $\mathrm{spec}(\tbP) \cap \mathrm{spec}(\tbR) = \varnothing$.
\begin{enumerate}\itemsep=0pt
\item[$(1)$] $\tilde{g}$ given by \eqref{CM_sol} satisfies
\begin{gather}
    (\gamma \tilde{g})^2 = I \qquad \mbox{and} \qquad
 \mathrm{tr}(\gamma \tilde{g}) = \mathrm{tr}(\gamma g_0) - 2   \mathrm{tr}(\bGamma)   .
       \label{red_conditions_1}
\end{gather}
\item[$(2)$] If moreover the relations
\begin{gather}
     \tbR^\dagger = \bGamma \tbP \bGamma   , \qquad
    \bU^\dagger = \bV  g_0   , \qquad
     g_0^\dagger = g_0  , \qquad
     \bGamma^\dagger = \bGamma
                   \label{dagger_cond}
\end{gather}
hold, then $\tilde{g}$ given by \eqref{CM_sol} is Hermitian.
\end{enumerate}
\end{proposition}

\begin{remark}
\label{rem:superposition}
Let the matrix data $(\tbP_i, \tbR_i, \bU_i, \bV_i, \bGamma_i)$ satisfy $\bGamma_i^2 = \bI_{n_i}$ and
\begin{gather*}
     (\bGamma_i   \tbP_i)^2 = -\bI_{n_i}   , \qquad
     (\bGamma_i   \tbR_i)^2 = -\bI_{n_i}   , \qquad
     g_0 \gamma   \bU_i = \bU_i   \bGamma_i   , \qquad
     \bGamma_i   \bV_i = \bV_i   g_0 \gamma     .
\end{gather*}
Set $\tbP = \mbox{block-diag}(\tbP_1, \ldots, \tbP_N)$,
$\tbR = \mbox{block-diag}(\tbR_1, \ldots, \tbR_N)$,
$\bGamma = \mbox{block-diag}(\bGamma_1, \ldots, \bGamma_N)$, and
\[
    \bU = (\bU_1, \ldots, \bU_N)  , \qquad
    \bV = \begin{pmatrix} \bV_1 \\ \vdots \\ \bV_N \end{pmatrix} .
\]
Then we have $\bGamma^2 = \bI$ and (\ref{Gg_constraints}) holds. If
$\mathrm{spec}(\tbP) \cap \mathrm{spec}(\tbR) = \varnothing$, the corresponding Sylvester equation
has a unique solution $\tbX$. According to part (1) of Proposition~\ref{prop:CMred_quadr_constr},
$\tilde{g}$ given by~(\ref{CM_sol}) satisf\/ies the reduction conditions~(\ref{red_conditions_1}).
This is a way to \emph{superpose} solutions from the class obtained in Section~\ref{sec:sol_gen},
preserving the constraints~(\ref{Gg_constraints}). We simply block-diagonally compose the matrix data
associated with the constituents. In an obvious way, this method can be extended to part~(2) of
Proposition~\ref{prop:CMred_quadr_constr}.
\end{remark}

\begin{remark}
\label{rem:mixing}
Let $n=2N$ and
\begin{gather*}
    \tbP = \begin{pmatrix} \check{\bP} & 0 \\
                                      0 & -\check{\bP}^{-1} \\
                  \end{pmatrix}   , \qquad
    \tbR = \begin{pmatrix} \check{\bR} & 0 \\
                                      0 & -\check{\bR}^{-1}
                  \end{pmatrix}   , \qquad
    \bGamma = \begin{pmatrix} 0 & \imag   \bI_N \\
                                      -\imag   \bI_N & 0
                  \end{pmatrix}   ,
\end{gather*}
where $\check{\bP}$ and $\check{\bR}$ are invertible block-diagonal $N \times N$ matrices, composed
of blocks of the form~(\ref{P-block_from_B-Jordan}). Then we have $(\bGamma \tbP)^2 = -\bI_n$ and
$(\bGamma \tbR)^2 = -\bI$. Choosing $\gamma$ and $g_0$ such that $(g_0 \gamma)^2=I$,
the conditions $g_0 \gamma \bU = \bU \bGamma$ and $\bGamma \bV = \bV g_0 \gamma $ are
solved by
\[
    \bU = \begin{pmatrix} \check{\bU} & \imag \, g_0 \gamma   \check{\bU} \\
                  \end{pmatrix}   , \qquad
    \bV = \begin{pmatrix} \check{\bV} \\
                                  -\imag   \check{\bV}   g_0 \gamma
                  \end{pmatrix}   ,
\]
where $\check{\bU}$ and $\check{\bV}$ are arbitrary constant $m \times N$, respectively
$N \times m$ matrices.
Writing
\[
     \tbX = \begin{pmatrix}
                       \check{\bX} & \check{\bR}^{-1} \check{\bZ} \check{\bP} \\
                       \check{\bZ} & \check{\bR} \check{\bX} \check{\bP}
                  \end{pmatrix} ,
\]
reduces the $2N \times 2N$ Sylvester equation~(\ref{CM_Sylv}) to the two
$N \times N$ Sylvester equations
\begin{gather}
   \check{\bX} \check{\bP} - \check{\bR} \check{\bX} = \check{\bV} \check{\bU}   , \qquad
   \check{\bZ} \check{\bP} + \check{\bR}^{-1} \check{\bZ}
     = - \imag   \check{\bV}   g_0 \gamma \check{\bU}   .
      \label{Sylv_eqs_check}
\end{gather}
If $\check{\bX}$ and $\check{\bZ}$ are invertibel, then $\tbR \tbX$ is
invertible\footnote{$(\tbR \tbX)^{-1}$ can be computed as a $2 \times 2$ block matrix.
The problem of evaluating the original expression for $\tilde{g}$ that involves $2N \times 2N$
matrices then reduces to that of evaluating only $N \times N$ matrix expressions.}.
Proposition~\ref{prop:CMred_quadr_constr}, part~(1), implies that~(\ref{CM_sol}) with the above matrix data
satisf\/ies $(\gamma \tilde{g})^2 = I$ and $\mathrm{tr}(\gamma \tilde{g}) = \mathrm{tr}(\gamma g_0)$.
With a suitable choice of $\gamma$ and $g_0$ we can achieve that (\ref{red_cond2}) holds.
To fulf\/il the remaining Hermiticity condition, one possibility is via part (2) of
Proposition~\ref{prop:CMred_quadr_constr}.
See also Examples~\ref{ex:Ernst_P-Jordan_R-diag} and~\ref{ex:EM_mixing}. Such solutions can be superposed
in the way described in Remark~\ref{rem:superposition}.

In the special case where $\check{\bR} = r \, \bI_N$,
the solutions of the Sylvester equations (\ref{Sylv_eqs_check}) are
$\check{\bX} = \check{\bV} \check{\bU}   (\check{\bP} - r   \bI_N)^{-1}$ and
$\check{\bZ} = - \imag   \check{\bV}   g_0 \gamma \check{\bU}  (\check{\bP} + r^{-1}   \bI_N)^{-1}$.
These expressions are not invertible if $N>m$, so in this particular case our solution
formula only works for $N \leq m$ (also see Example~\ref{ex:Ernst_P-Jordan_R-diag}).
\end{remark}

\begin{remark}
\label{rem:Harrison-transf}
The following observation in particular underlies the Harrison transformation \cite{Harr68,Kinn73}
which we consider in Example~\ref{ex:Harrison} below.
Let $H$ be an $m \times m$-matrix that satisf\/ies
\begin{gather}
    H^\dagger   \gamma  H = \gamma    .   \label{H_cond1}
\end{gather}
If $\tilde{g}$ satisf\/ies
\[
    (\gamma   \tilde{g})^2 = I   , \qquad
    \tilde{g}^\dagger = \tilde{g}   ,
\]
then also
\[
     \tilde{g}' = H \tilde{g} H^\dagger  ,
\]
and we have $\mathrm{tr}(\gamma   \tilde{g}') = \mathrm{tr}(\gamma   \tilde{g})$.
If $\tilde{g}$ has the form (\ref{CM_sol}) with Hermitian $g_0$ and $(\gamma  g_0)^2 =I$,
and if~$H$ also satisf\/ies
\begin{gather}
    H   g_0   H^\dagger = g_0   ,   \label{H_cond2}
\end{gather}
then the ef\/fect of the transformation
$\tilde{g} \mapsto \tilde{g}'$ amounts to the replacement
\begin{gather*}
    \bU \mapsto \bU' = H   \bU   , \qquad \bV \mapsto \bV' = \bV   H^{-1}   ,
\end{gather*}
which leaves the Sylvester equation (\ref{CM_Sylv}) invariant.
\end{remark}

\subsection{Solutions of the Ernst equation of general relativity}

We choose $m=2$ and
\begin{gather*}
     \gamma = \begin{pmatrix} 0 & \imag \\
                                -\imag & 0 \end{pmatrix} ,
\end{gather*}
and write
\[
     v = \begin{pmatrix} 1 \\
                                 \imag   \cE \end{pmatrix} ,
\]
with a complex function $\cE$ and its complex conjugate $\bar{\cE}$. Then (\ref{g_param})
takes the form
\[
 \tilde{g} = \frac{2}{\cE + \bar{\cE}} \begin{pmatrix} 1 & \frac{\imag}{2} (\cE - \bar{\cE}) \\
              \frac{\imag}{2} (\cE - \bar{\cE}) & \bar{\cE} \cE \end{pmatrix},
\]
so that
\[
     \cE = \frac{1-\imag   \tilde{g}_{21}}{\tilde{g}_{11}}  .
\]
(\ref{CM}) now becomes the \emph{Ernst equation}
\[
      (\mathrm{Re}\, \cE)  \big(\pa_\rho^2 + \rho^{-1} \pa_\rho + \pa_z^2\big)   \cE
    = (\cE_\rho)^2 + (\cE_z)^2   ,
\]
where e.g.~$\pa_\rho$ denotes the partial derivative with respect to~$\rho$.
This equation determines solutions of the stationary axially symmetric Einstein vacuum equations.
The following statements are easily verif\/ied.
\begin{enumerate}\itemsep=0pt
\item
Excluding $\tilde{g} = \pm \gamma$, the reduction conditions (\ref{red_cond1})
are equivalent to
\begin{gather}
  \tilde{g} \; \mbox{ real} , \qquad
  \det(\tilde{g})=1   , \qquad
  \tilde{g}_{12} = \tilde{g}_{21} . \label{red_m=2}
\end{gather}
(\ref{red_cond2}) is then automatically satisf\/ied.
\item
For real $\tilde{g}$,
the second of the reduction conditions (\ref{red_cond1}) implies the f\/irst.
As a consequence, Proposition~\ref{prop:CMred_quadr_constr}, part~(1), already
generates solutions of the Ernst equation.
\end{enumerate}
We will use these observations in the following examples.

\begin{example}[Kerr-NUT]
\label{ex:Kerr-NUT}
For the solution of the non-autonomous chiral model given in Example~\ref{ex:CM_diag} with $n=2$,
we have (also see Remark~\ref{rem:CM_sol_det})
\[
       \det(\tilde{g}) = \frac{p_1 p_2}{r_1 r_2}  \det(g_0)   ,
\]
with $p_i$, $r_i$ given by (\ref{p_i,r_i}). Choosing
\[
       g_0 = I_2 \, ,
\]
so that $\det(g_0)=1$, the second of the reduction conditions (\ref{red_m=2}) is solved by setting
\[
     p_2 = -\frac{1}{p_1}   , \qquad
     r_2 = -\frac{1}{r_1}   ,
\]
noting that $-1/p_i$ is given by the expression for $p_i$ with $j_i$ exchanged by $-j_i$.
We shall write~$p$,~$r$ instead of $p_1$, $r_1$.
With $\bU =(u_{ij})$, $\bV=(v_{ij})$, the remaining constraint $\tilde{g}_{12} = \tilde{g}_{21}$
is solved by\footnote{Another
solution is $u_{22} = u_{12} u_{21}/u_{11}, v_{22} = v_{12} v_{21}/v_{11}$. But this
leads to $ \tilde{g} = g_0$.}
\[
     u_{22} = - u_{11} u_{12}/u_{21}   , \qquad
     v_{22} = - v_{11} v_{21}/v_{12}   .
\]
In the following we assume that $u_{11}$ and $v_{11}$ are dif\/ferent from zero and write
\[
     u_{21} = u   u_{11}   , \qquad v_{12} = v   v_{11}   .
\]
Then $u_{11}$, $u_{12}$, $v_{11}$, $v_{21}$ drop out of $\tilde{g}$. Without restriction of
generality, we can therefore choose them as $u_{11}=1$, $u_{12}=-u$, $v_{11}=1$ and $v_{21}=-v$,
hence
\[
     \bU = \begin{pmatrix} 1 & -u \\
                                    u & 1 \end{pmatrix}   , \qquad
     \bV = \begin{pmatrix} 1 & v \\
                                   -v & 1 \end{pmatrix}  .
\]
Then $\bU$ and $\bV$ commute with $\gamma$.
$\tilde{g}$ is \emph{real} in particular if either of the following conditions is fulf\/illed.
\begin{enumerate}\itemsep=0pt
\item[(1)] $p$, $r$ real (which means $b$, $b'$ real) and $u$, $v$ real.
\item[(2)] $\bar{r} = -\frac{1}{p}$ (which means $\bar{b}'=b$ and $j=-j' \in \{\pm1\}$) and $v = \bar{u}$.
\end{enumerate}
The Ernst potential takes the form
\[
  \cE = \frac{(1+uv)   \frac{p+r}{p-r} - \imag   (u-v)   \frac{pr-1}{pr+1} + (u-\imag)(v-\imag)}
        {(1+uv)   \frac{p+r}{p-r} - \imag   (u-v)  \frac{pr-1}{pr+1} - (u-\imag)(v-\imag)}
         .
\]
By a shift of the origin of the coordinate $z$, we can arrange in both cases that
\begin{gather}
   p = \rho^{-1}   \left( z + b + j   \mathfrak{r}_+ \right), \qquad
   r = \rho^{-1}   \left( z - b + j'   \mathfrak{r}_- \right),\nonumber\\
   \mathfrak{r}_\pm := \sqrt{(z \pm b)^2 + \rho^2}  , \qquad  j,j' \in \{\pm1\},
         \label{p,r_expr}
\end{gather}
where $b \in \bbR$ in case (1) and $b \in \imag \, \bbR$ in case (2). Using
\begin{gather}
 \frac{p+r}{p-r} = \frac{1}{2b}   (j   \mathfrak{r}_+ + j'   \mathfrak{r}_- )   , \qquad
 \frac{pr-1}{pr+1} = \frac{1}{2b}   (j   \mathfrak{r}_+ - j'   \mathfrak{r}_- )    ,
             \label{p,r->r+-}
\end{gather}
and introducing
\begin{gather*}
     \mathfrak{a} = -b   \frac{1+ uv}{u-v}   , \qquad
     \mathfrak{l} = j   b   \frac{1-uv}{u-v}   , \qquad
     \mathfrak{m} = - j   b   \frac{u+v}{u-v}
   ,   
\end{gather*}
we obtain
\begin{gather}
  \cE = \frac{\mathfrak{r}_+ - j j'   \mathfrak{r}_-
        - \imag   \frac{\mathfrak{a}}{b}   ( \mathfrak{r}_+ + j j'   \mathfrak{r}_- )
        - 2   (\mathfrak{m} + \imag   \mathfrak{l})}{ \mathfrak{r}_+ - j j'   \mathfrak{r}_-
        - \imag   \frac{\mathfrak{a}}{b}  ( \mathfrak{r}_+ + j j'   \mathfrak{r}_- )
        + 2   (\mathfrak{m} + \imag   \mathfrak{l}) }    .   \label{Ernst_Kerr-NUT}
\end{gather}
Setting $j j'=-1$, the cases (1)
and (2) simply distinguish the \emph{non-extreme} and the \emph{hyperextreme}
\emph{Kerr-NUT} space-times (see e.g.~\cite{AGM01}).
The constants satisfy $\mathfrak{m}^2 + \mathfrak{l}^2 - \mathfrak{a}^2 = b^2$.
\end{example}

\begin{example}
\label{ex:Ernst_P-Jordan_R-diag}
Let $n=2N$. With the choices made in Remark~\ref{rem:mixing},
Proposition~\ref{prop:CMred_quadr_constr}, part (1), implies that $\tilde{g}$, given by (\ref{CM_sol})
with $g_0=I_2$, satisf\/ies $(\gamma \tilde{g})^2 = I_2$ and $\mathrm{tr}(\gamma \tilde{g}) =0$.
Choosing all parameters real, it follows that $\tilde{g} = I_2 + \bU (\tbR \tbX)^{-1} \bV$
determines a solution of the Ernst equation, provided that $\tbX$ is invertible\footnote{The latter
condition may indeed be violated, as shown in Remark~\ref{rem:mixing}.}.
For $N=1$, we are back to the preceding example. For $N=2$ let, for example,
\begin{gather*}
    \tbP = \begin{pmatrix} p & p   \mathfrak{r}^{-1} & 0 & 0 \\
                                      0 & p     & 0 & 0 \\
                                      0 & 0     & -p^{-1} & (p \mathfrak{r})^{-1} \\
                                      0 & 0     &         & -p^{-1}
                  \end{pmatrix}, \qquad
    \tbR = \begin{pmatrix} r_1 & 0 & 0 & 0 \\
                                      0 & r_2 & 0 & 0 \\
                                      0 & 0 & -r_1^{-1} & 0 \\
                                      0 & 0 &  0      & -r_2^{-1}
                  \end{pmatrix}, \\
    \bGamma = \begin{pmatrix}  0 & 0 & \imag & 0 \\
                                      0 & 0 & 0 & \imag \\
                                      -\imag & 0 & 0 & 0 \\
                                      0 & -\imag & 0 & 0
                  \end{pmatrix},
\end{gather*}
where $p = \rho^{-1}(z+b+\mathfrak{r})$, $\mathfrak{r}=\pm\sqrt{(z+b)^2+\rho^2}$, and $r_i$ is also
of the form (\ref{p_i,r_i}) with a constant $b_i' \neq b$. The conditions for $\bU$ and $\bV$ restrict
these matrices to the form
\[
    \bU = \begin{pmatrix} u_1 & u_3 & -u_2 & -u_4 \\
                                     u_2 & u_4 & u_1  & u_3
                  \end{pmatrix} , \qquad
    \bV = \begin{pmatrix} v_1 & v_2 \\
                                   v_3 & v_4 \\
                                  -v_2 & v_1 \\
                                  -v_4 & v_3
                  \end{pmatrix} ,
\]
with constants $u_i$, $v_i$. If $r_1=r_2 =:r$ (i.e.~$b_1'=b_2'$), it turns out that $\tilde{g}$ does
\emph{not} depend on~$r$ and~$v_i$, and we obtain
\[
     \cE = \frac{\mathfrak{r} + \imag \, \mathfrak{a}   (z+b)   \mathfrak{r}^{-1}
           - (\mathfrak{m}+\imag   \mathfrak{l})}
             {\mathfrak{r} + \imag   \mathfrak{a}   (z+b)   \mathfrak{r}^{-1}
           + (\mathfrak{m}+\imag   \mathfrak{l})}   ,
\]
with the parameters
\[
     \mathfrak{a} = \frac{u_1^2 + u_2^2}{2(u_1 u_4 - u_2 u_3)}   , \qquad
     \mathfrak{l} = \frac{u_1^2 - u_2^2}{2(u_1 u_4 - u_2 u_3)}   , \qquad
     \mathfrak{m} = - \frac{u_1 u_2}{u_1 u_4 - u_2 u_3}   ,
\]
which satisfy $\mathfrak{m}^2 - \mathfrak{a}^2 + \mathfrak{l}^2 =0$.
$\cE$ is the Ernst potential of an \emph{extreme} Kerr-NUT space-time.
\end{example}

\begin{example}[Multi-Kerr-NUT]
\label{ex:multi_Kerr-NUT}
According to Remark~\ref{rem:superposition}, there is a simple way to superpose solutions
by block-diagonally composing their matrix data.
Let
\begin{gather*}
    \bU_i = \begin{pmatrix} 1 & -u_i \\
                                 u_i & 1 \end{pmatrix} , \qquad
    \bV_i = \begin{pmatrix} 1 & v_i \\
                                 -v_i & 1 \end{pmatrix}  , \\
    \tbP_i = \begin{pmatrix} p_i & 0 \\
                                 0 & -1/p_i \end{pmatrix} , \qquad
    \tbR_i = \begin{pmatrix} r_i & 0 \\
                                 0 & -1/r_i \end{pmatrix} ,
\end{gather*}
where $p_i \neq r_k$, $i,k=1,\ldots,N$, are given by (\ref{p_i,r_i}), and either
$b_i,b_i',u_i,v_i \in \bbR$ or
$\bar{b}_i' = b_i \in \bbC$, $j_i'=-j_i$, $v_i=\bar{u}_i \in \bbC$ (cf.\ Example~\ref{ex:Kerr-NUT}).
Set
\[
    \bU = (\bU_1, \ldots, \bU_N)   , \qquad
    \bV = \begin{pmatrix}  \bV_1 \\ \vdots \\ \bV_N \end{pmatrix},
\]
and $\tbP = \mbox{block-diag}(\tbP_1, \ldots, \tbP_N)$,
$\tbR = \mbox{block-diag}(\tbR_1, \ldots, \tbR_N)$,
$\bGamma = \mbox{block-diag}(\gamma, \ldots, \gamma)$.
With $g_0 = I_2$, all assumptions of part (1) of Proposition~\ref{prop:CMred_quadr_constr} hold,
hence with these data~(\ref{CM_sol}) determines a family of solutions of the
Ernst equation. Obviously, such a solution is a superposition of $N$ (non-extreme, respectively
hyperextreme) Kerr-NUT solutions\footnote{See e.g.~\cite{Beli+Sakh79,Neug80,Kramer+Neug80}
for other derivations,
and also \cite{Neug+Hennig09}, as well as the references cited there.}.
More generally, in the same way we can superpose any number of solutions with matrix data of
the form given in Example~\ref{ex:Ernst_P-Jordan_R-diag}.
\end{example}

\subsection[Solutions of the Ernst equations in the Einstein-Maxwell case]{Solutions of the Ernst equations in the Einstein--Maxwell case}
Choosing
\begin{gather}
     \gamma = \begin{pmatrix} 0 & \imag & 0 \\
                                 -\imag & 0 & 0 \\
                                  0 & 0 & -I_{m-2} \end{pmatrix}  ,
                          \label{EM_gamma}
\end{gather}
and writing
\[
     v = \begin{pmatrix} 1 \\
                                \imag   \cE \\
                                \sqrt{2}   \Phi \end{pmatrix} ,
\]
with a complex function $\cE$ and a complex $(m-2)$-component vector $\Phi$, (\ref{g_param}) takes the form
\begin{gather*}
 \tilde{g} = \frac{2}{\cE \!+\! \bar{\cE} \!+\! 2   \Phi^\dagger \Phi}
 \begin{pmatrix}
  1 & \frac{\imag}{2} (\cE - \bar{\cE} + 2   \Phi^\dagger \Phi) & \sqrt{2}   \Phi^\dagger \\
  \frac{\imag}{2} (\cE - \bar{\cE} - 2   \Phi^\dagger \Phi) & \bar{\cE} \cE
               & \imag   \sqrt{2}   \cE   \Phi^\dagger \\
  \sqrt{2}   \Phi & -\imag   \sqrt{2}   \bar{\cE}  \Phi
        & 2 \Phi \Phi^\dagger - \frac{1}{2} (\cE + \bar{\cE} + 2   \Phi^\dagger \Phi) I_{m-2}
         \end{pmatrix}
\end{gather*}
(also see \cite{Guers+Xanth82,Guerses84}). We have
\[
    \cE = \frac{1 - \imag   \tilde{g}_{21}}{\tilde{g}_{11}}   , \qquad
    \Phi^\intercal = \frac{1}{\sqrt{2}   \tilde{g}_{11}}   (\tilde{g}_{31},\ldots,\tilde{g}_{m-2,1})
    ,
\]
where ${}^\intercal$ denotes transposition.
In the following we consider the case $m=3$, where (\ref{CM}) becomes the system of \emph{Ernst equations}
\begin{gather*}
      (\mathrm{Re}\, \cE + \bar{\Phi} \Phi)   \big(\pa_\rho^2 + \rho^{-1} \pa_\rho + \pa_z^2\big)   \cE
   =   (\cE_\rho)^2 + (\cE_z)^2 + 2   \bar{\Phi}   [ \Phi_\rho   \cE_\rho
      + \Phi_z   \cE_z ]    ,  \\
      (\mathrm{Re} \, \cE + \bar{\Phi} \Phi) \big(\pa_\rho^2 + \rho^{-1} \pa_\rho + \pa_z^2\big)   \Phi
   =  \cE_\rho   \Phi_\rho + \cE_z   \Phi_z
      + 2   \bar{\Phi}   \big[ (\Phi_\rho)^2 + (\Phi_z)^2 \big]   ,
\end{gather*}
which determine solutions of the stationary axially symmetric Einstein--Maxwell equations
(without further matter f\/ields). If $\cE =1$ and  $\Phi=0$, then $\tilde{g}$ reduces to
\begin{gather}
  g_0 = \begin{pmatrix} 1 & 0 & 0 \\
                                  0 & 1 & 0 \\
                                  0 & 0 & -1 \end{pmatrix} , \label{DN_g0}
\end{gather}
which corresponds to the Minkowski metric.

\begin{example}[Demia\'{n}ski--Newman]
Let $n=2$ and
\begin{gather*}
   \bGamma = \begin{pmatrix} 0 & \imag \\
                                -\imag & 0 \end{pmatrix} , \qquad
   \tbP = \begin{pmatrix} p & 0 \\
                                   0 & -1/p \end{pmatrix} , \qquad
   \tbR = \begin{pmatrix} r & 0 \\
                                   0 & -1/r \end{pmatrix},   
\end{gather*}
with $p$, $r$ as in (\ref{p_i,r_i}) with constants $b$, $b'$.
Solving $g_0 \gamma   \bU = \bU \bGamma$ and $\bGamma \bV = \bV g_0 \gamma$, and
recalling that $\bU$ and $\bV$ enter the solution formula (\ref{CM_sol})
only up to an overall factor, leads to
\begin{gather}
    \bU = \begin{pmatrix} 1 & -u  \\
                                   u  & 1\\
                                   s  & \imag   s  \end{pmatrix} , \qquad
    \bV = \begin{pmatrix}
                   1 & v & -t \\
                 -v  & 1 & \imag   t \end{pmatrix}.
        \label{neDN_UV_general}
\end{gather}
According to Proposition~\ref{prop:CMred_quadr_constr}, part (1), in order to obtain solutions
of the Ernst equations it remains to determine conditions under which $\tilde{g}$ is Hermitian.
By explicit evaluation one f\/inds that this is so if one of the following sets of conditions
is satisf\/ied.
\begin{enumerate}\itemsep=0pt
\item[(1)] $b,b' \in \mathbb{R}$ and
\begin{gather}
     s t = -2   \frac{v+\imag}{\bar{u}+\imag}  \,  \mathrm{Im} \, u    , \qquad
     |v+\imag|^2 \, \mathrm{Im}\,  u + |u-\imag|^2 \, \mathrm{Im}\,  v = 0
                       , \qquad
     2 \, \mathrm{Im} \, u +|s|^2 = 0   .               \label{DN_UVconds}
\end{gather}
\item[(2)] $\bar{b}' = b$, $j'=-j$, $v= \bar{u}$ and $t=\bar{s}$.
\end{enumerate}
Without restriction of generality we can set $b' = -b$, so that $p$ and $r$ are given by (\ref{p,r_expr}).
The Ernst potential $\cE$ is again of the form (\ref{Ernst_Kerr-NUT}), where now
\[
    \mathfrak{a} = -b   \frac{1 + u v - s t}{u-v + \imag   s t}   , \qquad
    \mathfrak{l} = j   b   \frac{1-u v}{u-v + \imag   s t}   , \qquad
    \mathfrak{m} = -j   b   \frac{u + v}{u-v + \imag   s t}   .
\]
The second Ernst potential is given by
\begin{gather}
    \Phi =  \frac{2   (q_e + \imag   q_m)}{ \mathfrak{r}_+ - j j'   \mathfrak{r}_-
        - \imag   \frac{\mathfrak{a}}{b}   ( \mathfrak{r}_+ + j j'   \mathfrak{r}_- )
        + 2   (\mathfrak{m} + \imag   \mathfrak{l}) }    ,
          \label{Ernst_Phi}
\end{gather}
where
\[
    q_e = - \frac{j   b}{\sqrt{2}}   \frac{s   (v-\imag) + t   (u + \imag)}{u-v + \imag   s t}
           , \qquad
    q_m = \imag   \frac{j   b}{\sqrt{2}}   \frac{s   (v-\imag) - t   (u+\imag)}{u-v+ \imag   s t}
           .
\]
In both cases, the parameters $\mathfrak{a}$, $\mathfrak{l}$, $\mathfrak{m}$, $q_e$, $q_m$
are real and satisfy
\begin{gather*}
    \mathfrak{m}^2 - \mathfrak{a}^2 + \mathfrak{l}^2 - q_e^2 - q_m^2 = b^2   . 
\end{gather*}
Cases (1) and (2) correspond to a \emph{non-extreme}, respectively \emph{hyperextreme},
\emph{Demia\'{n}ski--Newman} space-time (see e.g.~\cite{AGM01}).
$q_e$ and $q_m$ are the electric and magnetic charge, respectively.
Whe\-reas~(2) can be neatly expressed via~(\ref{dagger_cond}), we
have been unable so far to f\/ind a corresponding formulation for the conditions (1)
in terms of the matrices $\tbP$, $\tbR$, $\bU$, $\bV$, also see Remark~\ref{rem:multi-non-extreme}.
\end{example}

\begin{example}[Harrison transformation]
\label{ex:Harrison}
We can generate solutions of (\ref{DN_UVconds}) via a \emph{Harrison transformation}
\cite{Harr68,Kinn73}.
A non-extreme Kerr-NUT solution (without charge) corresponds to the data
\begin{gather*}
    \bU_{0} = \begin{pmatrix} 1 & -u_0  \\
                                   u_0  & 1\\
                                   0  & 0  \end{pmatrix} , \qquad
    \bV_{0} = \begin{pmatrix}
                   1 & v_0 & 0 \\
                 -v_0  & 1 & 0 \end{pmatrix},
\end{gather*}
with \emph{real} $u_0$, $v_0$.  The matrix
\begin{gather*}
    H = \frac{1}{1-|c|^2} \begin{pmatrix}
                   1 & \imag   |c|^2 & \imag   \sqrt{2}   c \\
                 -\imag   |c|^2  & 1 & \sqrt{2}   c\\
                 \imag   \sqrt{2}   \bar{c} & -\sqrt{2}   \bar{c} & -1-|c|^2
                                 \end{pmatrix}
\end{gather*}
with $c \in \bbC$ satisf\/ies (\ref{H_cond1}) and (\ref{H_cond2}). Then
$\bU' = H   \bU_{0}$ and $\bV' = \bV_{0}   H$
satisfy $g_0 \gamma   \bU' = \bU' \bGamma$ and $\bGamma \bV' = \bV' g_0 \gamma$, since
$\bU_{0}$ and $\bV_{0}$ satisfy these conditions.
Without ef\/fect on the solution of the chiral model, we can rescale these matrices to
\begin{gather*}
 \bU  =  \frac{1}{1 + \imag   u_0   |c|^2} \begin{pmatrix}
                   1 + \imag   u_0   |c|^2 & - (u_0 - \imag   |c|^2) \\
                 u_0 - \imag   |c|^2 & 1 + \imag   u_0   |c|^2  \\
                 \sqrt{2}   (\imag -u_0)   \bar{c} &  \imag   \sqrt{2}   (\imag-u_0)   \bar{c}
                                 \end{pmatrix} , \\
 \bV  =  \frac{1}{1 - \imag   v_0   |c|^2} \begin{pmatrix}
                1 - \imag   v_0   |c|^2 & v_0 + \imag   |c|^2 & \sqrt{2}   (\imag+v_0)   c \\
       -(v_0 + \imag   |c|^2) & 1 - \imag   v_0   |c|^2 & -\imag   \sqrt{2}   (\imag + v_0)   c
                                \end{pmatrix}  ,
\end{gather*}
which have the form (\ref{neDN_UV_general}) and indeed satisfy (\ref{DN_UVconds}).
Using (\ref{p,r->r+-}), the resulting Ernst poten\-tials~$\cE$ and~$\Phi$ can be written in the form
(\ref{Ernst_Kerr-NUT}), respectively (\ref{Ernst_Phi}), where now
\begin{gather*}
     \mathfrak{a} = - b   \frac{1+u_0 v_0}{u_0-v_0}   , \qquad
     \mathfrak{l} = j   b   \frac{1-u_0 v_0}{u_0-v_0}   \frac{1+|c|^2}{1-|c|^2}   , \qquad
     \mathfrak{m} = -j   b   \frac{u_0 + v_0}{u_0-v_0}   \frac{1+|c|^2}{1-|c|^2}   ,
\end{gather*}
and
\begin{gather*}
  q_e = \frac{2 j b  [ (u_0v_0-1)  \, \mathrm{Re}\, c - (u_0+v_0)   \,\mathrm{Im}\, c ]
                 }{(1-|c|^2)(u_0-v_0)}
               , \\
  q_m = -\frac{2 j b   [ (u_0+v_0)  \, \mathrm{Re} \, c + (u_0v_0-1) \, \mathrm{Im} \, c ]
                 }{(1-|c|^2)(u_0-v_0)}     .
\end{gather*}
\end{example}

\begin{example}[hyperextreme multi-Demia\'{n}ski--Newman]
\label{ex:multi-hyperDN}
Let
\begin{gather*}
    \bGamma = \begin{pmatrix}
       0 & \imag &        &        &         \\
  -\imag &     0 &        &        &         \\
         &       & \ddots &        &         \\
         &       &        & 0      & \imag   \\
         &       &        & -\imag & 0
                     \end{pmatrix}  , \qquad
  \begin{array}{l}
       \tbP = \mbox{block-diag}(\tbP_1,\ldots,\tbP_N), \\ \\
       \tbR = \mbox{block-diag}(\tbR_1,\ldots,\tbR_N), \\ \\
       \bU = (\bU_1, \ldots, \bU_N),
  \end{array}
 \qquad
    \bV = \begin{pmatrix} \bV_1 \\ \vdots \\ \bV_N \end{pmatrix} ,
\end{gather*}
with
\begin{gather*}
   \tbP_i = \begin{pmatrix} p_i & 0 \\
                                   0 & -1/p_i \end{pmatrix} ,\qquad
   \tbR_i = \begin{pmatrix} r_i & 0 \\
                                   0 & -1/r_i \end{pmatrix} ,\\
    \bU_i = \begin{pmatrix} 1 & -u_i  \\
                                  u_i  & 1  \\
                                  s_i  & \imag   s_i  \end{pmatrix} ,\qquad
    \bV_i = \begin{pmatrix}
                   1 & \bar{u}_i & -\bar{s}_i \\
           -\bar{u}_i  &        1 & \imag   \bar{s}_i \end{pmatrix} ,
\end{gather*}
where $\bar{r}_i =-1/p_i$.
Then the conditions of Proposition~\ref{prop:CMred_quadr_constr} are obviously satisf\/ied
(with~$\gamma$ and~$g_0$ given by~(\ref{EM_gamma}), respectively (\ref{DN_g0})). It follows that
(\ref{CM_sol}) determines a solution of the (Einstein--Maxwell--) Ernst equations.
This is a superposition of $N$ hyperextreme Demia\'{n}ski--Newman solutions.
\end{example}

\begin{remark}
\label{rem:multi-non-extreme}
Whereas the hyperextreme \emph{multi}-Demia\'{n}ski--Newman solutions are obtained in a
straightforward way, this is not so in the non-extreme case. So far a suitable
condition on the matrix data is lacking. Similar problems are known in other
approaches, see e.g.~\cite[Section~34.8]{SKMHH03}.
\end{remark}

\begin{example}
\label{ex:EM_mixing}
Let $n=2N$. With the choices made in Remark~\ref{rem:mixing},
Proposition~\ref{prop:CMred_quadr_constr}, part (1), implies that the expression
given by~(\ref{CM_sol}) with $g_0$ in (\ref{DN_g0})
satisf\/ies $(\gamma \tilde{g})^2 = I_3$ and $\mathrm{tr}(\gamma \tilde{g}) =1$.
In order to obtain a solution of the Ernst equations, it suf\/f\/ices to arrange that $\tilde{g}$
is Hermitian. A suf\/f\/icient condition is given by part~(2) of Proposition~\ref{prop:CMred_quadr_constr}.
This leads to a huge family of solutions of the Einstein--Maxwell equations. The hyperextreme
Demia\'{n}ski--Newman solution is just the simplest example in this family. Furthermore,
such solutions can be superposed in the simple way described in Remark~\ref{rem:superposition}
and Example~\ref{ex:multi-hyperDN}.
An exploration of the corresponding space-times would be a dif\/f\/icult task.
\end{example}

\section{Conclusions}
\label{sec:conclusions}
We addressed the $m \times m$ non-autonomous chiral model in a new way, starting
from a very simple and universal solution generating result
within the bidif\/ferential calculus approach. This resulted in an inf\/inite family of
exact solutions for any matrix size $m$, parametrized by matrices subject to a Sylvester equation.
To solve the latter is a well-studied and fairly simple problem. At least in the compact
form presented in this work, according to our knowledge these solutions have not appeared
previously in the literature.

The non-autonomous chiral model originally appeared in reductions of Einstein's equations.
We demonstrated in Section~\ref{sec:Ernst} that the ``multi-solitons'' on a f\/lat background,
known in the case of stationarity, axial symmetry and vacuum, respectively electrovacuum, are indeed
contained in the family of solutions that we obtained in Section~\ref{sec:naCM} for the
non-autonomous chiral model equation.
More precisely, we found conditions to be imposed on the matrices that parametrize the latter solutions
such that (in the cases $m=2$, respectively $m=3$) they become solutions of the
Ernst equation(s) of general relativity. Only in the case of non-extreme multi-Demia\'{n}ski--Newman
solutions we were not (yet) able to f\/ind a corresponding characterization of the matrix data.

Beyond the solutions found e.g.\ by Belinski and Zakharov, which in the present work correspond
to diagonal matrices $\tbP$ and $\tbR$, there are solutions associated with non-diagonal
matrix data. It may well be that such solutions can be obtained alternatively e.g.\ in the
Belinski--Zakharov framework with a dressing matrix involving higher order poles, or by
taking suitable limits where some poles coincide. In any case, our approach yields these
solutions directly.
Moreover, relaxing the spectral condition for the matrices $\tbP$ and $\tbR$,
the Sylvester equation has further solutions, provided that the matrix $\bV \bU$ on its right hand
side is appropriately chosen. This is another possibility to obtain new solutions.
Finally, we should mention the possibility to make sense of the
limit\footnote{See e.g.~\cite{Bhat+Rose97,GHS06} for results on the operator Sylvester equation.}
$n \to \infty$ (where $n \times n$ is the size of the matrices that parametrize the solutions).
In conclusion, at present it is not quite clear what the generated class of solutions really embraces.

Moreover, using the original method of
Belinski and Zakharov, in the Einstein--Maxwell case no appropriate reduction conditions could
be found (cf.~\cite{Aleks10}), and a dif\/ferent approach had to be developed
\cite{Alek88,Beli+Verd01}. We had less problems in this respect.

On the other hand, the Belinski--Zakharov approach, the modif\/ied approach of Alekseev
\cite{Alek88} in the Einstein--Maxwell case, and others can also be used to generate
``solitons'' on a \emph{non-flat} background. Perhaps a corresponding
extension of Proposition~\ref{prop:main} exists. This is also suggested by the relation
with Darboux transformations in Appendix~\ref{app:gt}. In any case, here we have a~limitation
of Proposition~\ref{prop:main} (which is not a limitation of the bidif\/ferential calculus framework,
which of\/fers various methods~\cite{DMH08bidiff}),
but we have the advantage of a very simple and general result that covers physically interesting cases.

The appearance of a Sylvester equation is a generic feature of the solution generating
result formulated in Proposition~\ref{prop:main} and in Appendix~\ref{app:gt} (also see~\cite{DMH08bidiff}). Sylvester
equations and their simplest solutions, Cauchy-like matrices, frequently appeared in
the integrable systems literature. But this is the f\/irst
time we came across a Sylvester equation involving non-constant matrix data.
A particularly nice feature is the fact that solutions can be superposed by simply
composing their matrix data into bigger block-diagonal matrices.
The corresponding Sylvester equation still has to be solved, but a unique solution
exists if we impose a not very restrictive spectral condition on these matrix data.

In Appendix~\ref{app:CM_linsys} we recovered two familiar Lax pairs for the non-autonomous chiral
model from the general linear equation (\ref{bidiff_linsys}) in the bidif\/ferential calculus
framework. Our way toward exact solutions in Section~\ref{sec:naCM} is more closely related
to Maison's Lax pair than to that of Belinski and Zakharov. We eliminated the $\theta$-dependence,
whereas in the Lax pair of Belinski and Zakharov the $\theta$-dependence is kept and it
involves derivatives with respect to this ``spectral parameter''.

\looseness=-1
Our results extend beyond the Einstein--Maxwell case and are also applicable to
higher-dimensional gravity theories (see e.g.~\cite{Lee85,Lee87II,Wei01,Harm04,Pomer06,Azuma+Koikawa06,Iguchi+Mishima06,TMY06,Yazad08,Empa+Reall08,FJRV10}).
Besides that, other reductions of the non-autonomous chiral model (for some $m$) are
of interest, see e.g.~\cite{Mikh+Yare82,Guts+Lipo95}, and the set of solutions that we obtained
in this work will typically be reducible to solutions of~them.

Since Proposition~\ref{prop:main} actually generates solutions of a Miura transformation
equation, we obtained simultaneously solutions of the \emph{Miura-dual} of the non-autonomous
chiral model equation: under the assumptions of
Proposition~\ref{prop:CM_sol}, $\tilde{\phi} = \bU \tbX^{-1} \bV$ solves the
equation (\ref{CM_Miura_dual}). The further results in Section~\ref{sec:naCM},
in particular Example~\ref{ex:CM_diag}, then provide us with explicit families of solutions.

\looseness=-1
Proposition~\ref{prop:main}, respectively Theorem~\ref{theorem:main}, can
actually be formulated and proved without explicit use of the two \emph{non}linear
equations involving only $\bP$, respectively~$\bR$. In such a~formulation, the theorem generates
solutions of the nonlinear integrable equation (\ref{phi_eq}), respectively~(\ref{g_eq}),
from solutions of \emph{li\-near} equations. However, the equations for~$\bP$ and~$\bR$
arise as integrability conditions of the latter. In previous work
\cite{DMH08bidiff,DMH10NLS,DMH10AKNS,DKMH11}, we chose
$\bP$ and $\bR$ as $\d$- and $\bd$-\emph{constant} matrices, which indeed reduces the equations that
have to be solved to only \emph{li\-near} ones, and we recovered (and somewhat generalized)
known soliton solution families. In case of the non-autonomous chiral model and, more specif\/ically,
its reduction to the Ernst equation, it turned out to be necessary to go beyond this level,
and thus to consider genuine solutions of the nonlinear equations for $\bP$ and $\bR$, in
order to obtain relevant solutions like those associated with multi-Kerr-NUT space-times
and their (electrically and magnetically) charged generalizations.
This also suggests a corresponding application of the theorem (or Proposition~\ref{prop:main})
to other integrable PDDEs.



\appendix

\section{Via a Darboux transformation and a projection\\ to a
non-iterative solution generating result}
\label{app:gt}

\begin{lemma}
\label{lemma:Darboux}
Let $P$ be invertible. The transformation
\begin{gather}
    (\phi, g) \quad \mapsto \quad (\phi',g') = \big(\phi + X   P   X^{-1} ,
                    X   P   X^{-1}   g \big)   ,   \label{Darboux_transf}
\end{gather}
where $X$ is an invertible solution of \eqref{bidiff_linsys} with
$\bbA = \d \phi = (\bd g)   g^{-1}$, and $\bd P = (\d P)   P$,
maps a~solution of the Miura transformation equation \eqref{Miura}
into another solution.
\end{lemma}

\begin{proof}  Using (\ref{bidiff_linsys}) and $\bd P = (\d P)   P$, a direct computation leads to
\[
   (\bd g')   {g'}^{-1} - \d \phi'
 = \bbA - \d \phi - X   P   X^{-1} \big[\bbA -(\bd g)   g^{-1}\big]   X P^{-1} X^{-1}   ,
\]
which vanishes if $\bbA = \d \phi = (\bd g)   g^{-1}$.
\end{proof}

(\ref{Darboux_transf}) is an essential part of a \emph{Darboux transformation}, cf.~\cite{DMH08bidiff}.
In the following we will use this result to derive a theorem which
essentially reduces to Proposition~\ref{prop:main}, see Remark~\ref{rem:cor}.

\begin{lemma}\sloppy
\label{lemma:proj}
Let $(\bphi,\bg)$ be a solution of the Miura transformation equation \eqref{Miura}
in $\mathrm{Mat}(n,n,\cB)$.
Let $\bU \in \mathrm{Mat}(m,n,\cB)$ and $\bV \in \mathrm{Mat}(n,m,\cB)$
be $\d$- and $\bar{\d}$-constant. If
\begin{gather}
      \bphi = \bV \bU   \hat{\bphi}   ,  \label{bphi_VU}
\end{gather}
with some $\hat{\bphi} \in \mathrm{Mat}(n,n,\cB)$, then
\begin{gather}
    \phi = \bU \hat{\bphi} \bV   , \qquad
    g = \big(\bU \bg^{-1} \bV\big)^{-1}   \label{phi,g_via_proj}
\end{gather}
solve the Miura transformation equation \eqref{Miura}  in $\mathrm{Mat}(m,m,\cB)$.
\end{lemma}

\begin{proof} Since $(\bphi,\bg)$ is assumed to solve (\ref{Miura}), we have
\[
    \bd \bg^{-1} = - \bg^{-1}   \d \bphi = - \bg^{-1}   \bV \bU   \d \hat{\bphi}   .
\]
Multiplying by $\bU$ from the left and by $\bV$ from the right, we obtain
\[
    \bd g^{-1} = - g^{-1}   \d \phi   ,
\]
which is equivalent to (\ref{Miura}).
\end{proof}

\begin{theorem}\sloppy
\label{theorem:main}
Let $(-\bR,\bS)$ be a solution of the Miura transformation equation \eqref{Miura}  in
$\mathrm{Mat}(n,n,\cB)$, i.e.
\begin{gather}
     \bd \bS = - (\d \bR)   \bS   ,  \label{t1}
\end{gather}
and $\bS$ invertible.
Let $\bX$ be an invertible solution of the linear equation \eqref{bidiff_linsys}, now
in $\mathrm{Mat}(n,n,\cB)$ and with invertible $\bP$, hence
\begin{gather}
   \bar{\d} \bX = (\d \bX)   \bP - (\d \bR)   \bX    , \qquad
   \bar{\d} \bP = (\d \bP)   \bP   .  \label{t2}
\end{gather}
In addition we require that
\begin{gather}
    \bX   \bP - \bR   \bX = \bV   \bU   \bY    , \qquad
    \bd \bR = \bR   \d \bR   , \qquad
    \bar{\d} \bY = (\d \bY)   \bP   ,   \label{t3}
\end{gather}
where $\bU \in \mathrm{Mat}(m,n,\cB)$ and $\bV \in \mathrm{Mat}(n,m,\cB)$
are $\d$- and $\bar{\d}$-constant, and $\bY \in \mathrm{Mat}(n,n,\cB)$. Then also
\begin{gather}
    \phi = \bU \bY \bX^{-1} \bV   \qquad \mbox{and} \qquad
    g = (\bU \bS^{-1} \bX \bP^{-1} \bX^{-1} \bV)^{-1}   \label{th_solutions}
\end{gather}
solve the Miura transformation equation \eqref{Miura}, and thus \eqref{phi_eq},
respectively \eqref{g_eq}.
\end{theorem}

\begin{proof} Since we assume that $(-\bR,\bS)$ solves the Miura transformation equation
(\ref{Miura}) in $\mathrm{Mat}(n,n,\cB)$, according to Lemma~\ref{lemma:Darboux} this
also holds for the pair
\[
     \bphi = -\bR + \bX \bP \bX^{-1}   , \qquad
     \bg = \bX \bP \bX^{-1}   \bS   .
\]
Using the f\/irst of (\ref{t3}), we f\/ind that (\ref{bphi_VU}) holds with
$\hat{\bphi} = \bY   \bX^{-1}$. Now (\ref{phi,g_via_proj}) yields the
asserted formulas for $\phi$ and $g$. According to Lemma~\ref{lemma:proj}, $\phi$ and $g$
solve the Miura transformation equation~(\ref{Miura}).

Together with (\ref{t2}), the f\/irst of (\ref{t3}) implies
\[
    \bV \bU   ( \bd \bY - (\d \bY) \bP ) = (\bR   \d \bR - \bd \bR)   \bX   ,
\]
which is satisf\/ied if the last two conditions of (\ref{t3}) hold.
\end{proof}

\begin{remark}
This theorem generalizes a previous result in \cite{DMH08bidiff}, which has been applied
in \cite{DMH08bidiff,DMH10NLS,DMH10AKNS,DKMH11} with $\d$- and $\bd$-constant $\bP$, $\bR$,
in which case only \emph{linear} equations have to be solved
in order to generate solutions of~(\ref{phi_eq}), respectively~(\ref{g_eq}).

The above derivation shows that the theorem may be regarded as a combination of a
Darboux transformation (Lemma~\ref{lemma:Darboux}),
on the level of matrices of arbitrary size, and a projection mechanism (Lemma~\ref{lemma:proj}).
The projection idea can be traced back to work of Marchenko~\cite{March88}.
More generally, the above result can be formulated in
terms of suitable operators, replacing the matrices that involve a size $n$.
\end{remark}

The next remark shows that, with mild additional assumptions, Theorem~\ref{theorem:main}
reduces to Proposition~\ref{prop:main}.

\begin{remark}
\label{rem:cor}
The transformation
\[
    \bX \mapsto \bX   \bQ   , \qquad
    \bY \mapsto \bY   \bQ   , \qquad
    \bP \mapsto \bQ^{-1} \bP \bQ   ,
\]
with any invertible $\bQ \in \mathrm{Mat}(n,n,\cB)$, leaves
the expressions for $\phi$ and $g$ in (\ref{th_solutions}) invariant.
This is then also a symmetry transformation of (\ref{t2})
and (\ref{t3}) if $\bd \bQ^{-1} = (\d \bQ^{-1})   \bP$.
As a~consequence, under the assumptions that $\bY$ is invertible, without restriction
of generality we can set $\bY = \bI$, where $\bI$ is the $n \times n$ identity matrix.
Then $\phi$ is given by the expression in~(\ref{prop_phi,g}).

We further note that (\ref{t1}) and the second of (\ref{t3}) imply $\bd(\bR \bS) =0$.
Assuming that $\bR$ is invertible, we thus have $\bS = \bR^{-1} \bC$ with an
invertible $\bd$-constant $\bC$.
The expression for $g$ in~(\ref{th_solutions}) now takes the form
\[
    g = \big(\bU \bC^{-1} \bR   \bX (\bX \bP)^{-1} \bV\big)^{-1}
      = \big(I - \bU (\bX \bP)^{-1} \bV\big)^{-1}   \big(\bU \bC^{-1} \bV\big)^{-1}   ,
\]
assuming temporarily invertibility of $\bU \bC^{-1} \bV$.
Together with $\phi$, this remains a solution of (\ref{Miura}) if we drop the
last factor, so that
\[
    g = \big(I - \bU (\bX \bP)^{-1} \bV\big)^{-1}    .
\]
This expression also makes sense without the above additional invertibility assumptions.
We can still translate it into a simpler form. From the f\/irst of (\ref{t3}),
which now has the form of the last of (\ref{prop_conds}), we obtain
\[
    (\bR   \bX)^{-1} - (\bX   \bP)^{-1} = (\bR   \bX)^{-1} \bV   \bU   (\bX   \bP)^{-1}   .
\]
Multiplication by $\bU$ from the left and by $\bV$ from the right, and use in our last
formula for $g$, leads to the expression for $g$ in~(\ref{prop_phi,g}).
\end{remark}

\section{Linear systems for the non-autonomous chiral model}
\label{app:CM_linsys}

\subsection{Maison's Lax pair}
\label{subapp:CM_linsys}
Using the bidif\/ferential calculus determined by (\ref{CM_bidiff}), (\ref{bidiff_linsys})
with $\bbA = (\bd g)  g^{-1}$ takes the form
\begin{gather*}
    X_\rho + \rho^{-1} X_\theta = \big(g_\rho + \rho^{-1} g_\theta\big)   g^{-1} X
          - X_z   P   e^\theta   , \qquad
    X_z = g_z   g^{-1} X + \big(X_\rho - \rho^{-1} X_\theta\big)   P   e^\theta   ,
\end{gather*}
and (\ref{bidiff_P_eq}) reads
\begin{gather*}
    e^{-\theta}   \big(P_\rho + \rho^{-1} P_\theta\big) = - P_z   P   , \qquad
    P_z = e^\theta \big( P_\rho - \rho^{-1} P_\theta \big)   P   .
\end{gather*}
Disregarding a constant solution (cf.~Section~\ref{subapp:BZ}), we can eliminate
the $\theta$-dependence in the latter equations via
\[
    P = e^{-\theta}   \tilde{P}   ,
\]
with $\tilde{P}$ independent of $\theta$, and obtain
\begin{gather*}
   \tilde{P}_\rho = \rho^{-1} \tilde{P} \big (I - \tilde{P}^2\big) \big(I + \tilde{P}^2\big)^{-1}   , \qquad
    \tilde{P}_z = 2 \rho^{-1} \tilde{P}^2 \big(I + \tilde{P}^2\big)^{-1}   .
\end{gather*}
Furthermore, setting
\[
     X = e^{c_1 \theta}   \tilde{X}   , \qquad
     g = e^{c_2 \theta}   \tilde{g}   ,
\]
with $\tilde{X}$, $\tilde{g}$ independent of $\theta$,
the above linear system becomes
\begin{gather*}
    \tilde{X}_\rho \big(I + \tilde{P}^2\big)  =  \big(\tilde{g}_\rho
          + c_2   \rho^{-1} \tilde{g}\big)   \tilde{g}^{-1} \tilde{X}
          - \tilde{g}_z   \tilde{g}^{-1}   \tilde{X}   \tilde{P}
          - c_1   \rho^{-1} \tilde{X} \big(I - \tilde{P}^2\big)   , \\
    \tilde{X}_z   \big(I + \tilde{P}^2\big)  =  \tilde{g}_z   \tilde{g}^{-1} \tilde{X}
          + \big(\tilde{g}_\rho + c_2   \rho^{-1} \tilde{g}\big)   \tilde{g}^{-1} \tilde{X}   \tilde{P}
          - 2 c_1   \rho^{-1} \tilde{X}   \tilde{P}   .
\end{gather*}
Choosing
\[
      \tilde{P} = p   I
\]
with a function $p(\rho,z)$, the equations for $\tilde{P}$ can easily be integrated,
which results in\footnote{We note that $p$ is $\tbP_1$ in Section~\ref{subsec:CM_Jordan}.}
\[
      p = \rho^{-1}   \left( z + b \pm \sqrt{(z+b)^2 + \rho^2} \right)   ,
\]
where $b$ is an arbitrary constant. In terms of
\[
      \hat{X} = \tilde{g}^{-1} \tilde{X}
\]
the above linear system, simplif\/ied by setting $c_1=c_2=0$, then takes the form
\[
  \hat{X}_\rho = - \frac{p}{1+p^2} \big(\tilde{g}^{-1} \tilde{g}_z
                 + p   \tilde{g}^{-1} \tilde{g}_\rho\big)   \hat{X}   , \qquad
  \hat{X}_z = \frac{p}{1+p^2}   \big(\tilde{g}^{-1} \tilde{g}_\rho
              - p   \tilde{g}^{-1} \tilde{g}_z\big)   \hat{X}   .
\]
This system is equivalent to a linear system for the non-autonomous chiral model, f\/irst
found by Maison in 1979 \cite{Maison79} (also see~\cite{Kanning10}).

\subsection[The Belinski-Zakharov Lax pair]{The Belinski--Zakharov Lax pair}
\label{subapp:BZ}
Using instead of $\theta$ the variable
\[
     \lambda = - \rho   e^\theta   ,
\]
(\ref{CM_bidiff}) translates into
\begin{gather*}
    \d f = - f_z   \zeta_1 - \rho^{-1} \lambda   f_\rho   \zeta_2   , \qquad
    \bar{\d} f = - \big( \rho   \lambda^{-1} f_\rho + 2   f_\lambda \big)   \zeta_1
                 + f_z   \zeta_2   .  \label{CM_bidiff_BZ}
\end{gather*}
We consider the linear system (\ref{bidiff_linsys}) with $P=I$, which trivially solves (\ref{bidiff_P_eq}),
i.e.
\[
    \bd X = \bbA   X + \d X   .
\]
Writing
\begin{gather*}
      \bbA = -\frac{\rho}{\lambda}   A   \zeta_1 + B   \zeta_2   ,
\end{gather*}
the integrability condition (\ref{bidiff_linsys_integr}) takes the form
\[
     B_\rho - A_z = [A,B]   , \qquad (\rho A)_\rho + (\rho B)_z = 0   ,
\]
assuming that $A$, $B$ are $\lambda$-independent. Solving the f\/irst (zero curvature)
condition by
\[
     A = g_\rho   g^{-1}   , \qquad B = g_z   g^{-1}   ,
\]
the second equation becomes the non-autonomous chiral model equation
\[
    \big(\rho   g_\rho   g^{-1}\big)_\rho + \big(\rho   g_z   g^{-1}\big)_z = 0   .
\]
The above linear equation leads to
\[
    X_\rho = \frac{ \rho   \cU + \lambda   \cV }{\rho^2 + \lambda^2}    X
             - \frac{2 \rho \lambda}{\rho^2 + \lambda^2}   X_\lambda   , \qquad
    X_z = \frac{ \rho   \cV - \lambda   \cU }{\rho^2 + \lambda^2}    X
             + \frac{2 \lambda^2}{\rho^2 + \lambda^2}   X_\lambda   ,
\]
where
\[
    \cU = \rho   A   , \qquad
    \cV = \rho   B   .
\]
This is the Belinski--Zakharov Lax pair \cite{Beli+Sakh79} (also see~\cite[Chapter~8]{Beli+Verd01}). We note that the ``spectral parameter'' $\lambda$ has its origin in a
coordinate of the self-dual Yang--Mills equation. We also note that
$\bbA = (\bd g)   g^{-1}$ (using $g_\lambda =0$).

\section{Some proofs}
\label{app:proofs}

\subsection{Proof of Lemma~\ref{lemma:tbP-id}}
(1) Assuming that $\bI + \tbP^2$ is invertible, the system (\ref{tbP_eqs})
can be decoupled into
\begin{gather}
    \tbP_\rho = \rho^{-1}   \tbP   \big(\bI - \tbP^2\big)\big(\bI + \tbP^2\big)^{-1}   , \qquad
    \tbP_z = 2 \rho^{-1}  \tbP^2  \big (\bI + \tbP^2\big)^{-1}    ,
             \label{tbP_eqs_decoupled}
\end{gather}
which can also be written as
\[
     \big(\tbP^{-1}\big)_\rho = -\rho^{-1}   \tbP^{-1} \big(\bI - \tbP^2\big)\big(\bI + \tbP^2\big)^{-1}   , \qquad
     \big(\tbP^{-1}\big)_z = - 2 \rho^{-1} \big(\bI + \tbP^2\big)^{-1}   ,
\]
assuming that $\tbP$ is invertible. Subtraction yields
\[
    \big(\tbP - \tbP^{-1}\big)_\rho = - \rho^{-1} \big(\tbP - \tbP^{-1}\big)   , \qquad
    \big(\tbP - \tbP^{-1}\big)_z = 2 \rho^{-1}   \bI   ,
\]
which can be integrated to
\begin{gather}
    \tbP - \tbP^{-1} = 2 \rho^{-1}   (z   \bI + \bB)   ,  \label{tbP_id2}
\end{gather}
with a constant matrix $\bB$. This implies (\ref{tbP_id}).

(2) Let $\tbP$ satisfy (\ref{tbP_id}) with a constant matrix $\bB$. Then $\tbP$ is invertible,
since the existence of a non-vanishing vector annihilated by $\tbP$ would be in conf\/lict with
(\ref{tbP_id}). Thus (\ref{tbP_id2}) holds, which implies $[\tbP,\bB]=0$.
Dif\/ferentiation of (\ref{tbP_id}) with respect to $\rho$, and elimination of
$z \bI + \bB$ with the help of (\ref{tbP_id}) or equivalently (\ref{tbP_id2}), leads to
\begin{gather*}
   0 = \tbP_\rho   \tbP + \tbP \tbP_\rho + 2 \rho^{-2} (z \bI + \bB)   \tbP
         - 2 \rho^{-1} (z \bI + \bB)   \tbP_\rho
     = \big(\tbP + \tbP^{-1}\big) \tbP_\rho + \rho^{-1} \big(\tbP^2 - \bI\big)   ,
\end{gather*}
where we used the assumption $[\tbP_\rho,\tbP]=0$.
If $\bI + \tbP^2$ is invertible, the resulting equation is the f\/irst of (\ref{tbP_eqs_decoupled}).
In the same way we obtain the second of~(\ref{tbP_eqs_decoupled}).
(\ref{tbP_eqs_decoupled}) is equivalent to~(\ref{tbP_eqs}).

\subsection{Proof of Proposition~\ref{prop:CMred_quadr_constr}}

Using (\ref{CM_Sylv}), (\ref{Gg_constraints}) and (\ref{Gg0_conds}), we f\/ind
\[
     \tbR   (\tbX \tbP + \bGamma \tbR \tbX \bGamma) - (\tbX \tbP + \bGamma \tbR \tbX \bGamma)   \tbP = 0   ,
\]
so that the spectrum condition implies
\begin{gather}
    \bGamma \tbR \tbX \bGamma = - \tbX \tbP   .  \label{GammaRXGamma=XP}
\end{gather}
With the help of this result we obtain
\begin{gather*}
     g_0 \gamma \big(I + \bU (\tbR \tbX)^{-1} \bV \big)
  = g_0 \gamma + \bU \bGamma (\tbR \tbX)^{-1} \bV
 = g_0 \gamma - \bU (\bGamma \tbX \tbP)^{-1} \bV\\
 \hphantom{g_0 \gamma \big(I + \bU (\tbR \tbX)^{-1} \bV \big)}{}
 = \big(I - \bU (\tbX \tbP)^{-1} \bV\big)   g_0 \gamma   .
\end{gather*}
Using $( g_0 \gamma)^2 = I$, the condition $(\gamma \tilde{g})^2 = I$ for (\ref{CM_sol})
is therefore equivalent to
\[
    \big(I - \bU (\tbX \tbP)^{-1} \bV\big) \big(I + \bU (\tbR \tbX)^{-1} \bV \big) = I   .
\]
Expanding the left hand side and using the Sylvester equation (\ref{CM_Sylv})
to eliminate $\bV \bU$, this indeed turns out to be satisf\/ied.
To complete the proof of (1), it remains to derive the trace formula.
Using (\ref{CM_sol}), (\ref{CM_Sylv}) and (\ref{GammaRXGamma=XP}), we obtain
\begin{gather*}
     \mathrm{tr}(\gamma \tilde{g}) - \mathrm{tr}(\gamma g_0)
  =  \mathrm{tr}\big((\tbR \tbX)^{-1} \bV   g_0 \gamma \bU \big)
  = \mathrm{tr}\big( (\tbR \tbX)^{-1} \bV \bU \bGamma \big) \\
\phantom{\mathrm{tr}(\gamma \tilde{g}) - \mathrm{tr}(\gamma g_0)}{}
 =  \mathrm{tr}\big( (\tbR \tbX)^{-1} (\tbX \tbP - \tbR \tbX) \bGamma \big)
  = - \mathrm{tr}( \bGamma ) + \mathrm{tr}\big( (\tbR \tbX)^{-1} \tbX \tbP \bGamma\big) \\
\phantom{\mathrm{tr}(\gamma \tilde{g}) - \mathrm{tr}(\gamma g_0)}{}
  =  - \mathrm{tr}( \bGamma ) - \mathrm{tr}\big( (\tbR \tbX)^{-1} \bGamma \tbR \tbX\big)
  = - 2 \, \mathrm{tr}( \bGamma )   .
\end{gather*}

In order to prove (2), we consider the Hermitian conjugate of the Sylvester equation
(\ref{CM_Sylv}).
By use of (\ref{dagger_cond}), and with the help of
$g_0 \gamma   \bU = \bU   \bGamma$,
$\bGamma   \bV = \bV   g_0 \gamma$ and $(g_0 \gamma)^2=I$, it takes the form
\[
    \bV \bU = - \big(\bGamma \tbX^\dagger   \bGamma\big) \tbP
                  + \tbR   \big(\bGamma   \tbX^\dagger   \bGamma\big)   .
\]
By comparison with the original Sylvester equation, the spectrum condition allows us
to conclude that
\[
     \tbX^\dagger = - \bGamma \tbX   \bGamma   .
\]
Together with (\ref{GammaRXGamma=XP}) this implies
\[
     (\tbR \tbX)^\dagger
  = \tbX^\dagger \tbR^\dagger
  = - \bGamma \tbX   \bGamma^2 \tbP \bGamma
  = - \bGamma \tbX \tbP \bGamma
  = \tbR \tbX   .
\]
It follows that $\bU (\tbR \tbX)^{-1} \bV   g_0$ is Hermitian, and thus also~$\tilde{g}$
given by~(\ref{CM_sol}).

\subsection*{Acknowledgements}
We would like to thank Vladimir S.~Manko and 
anonymous referees for helpful comments.
During the course of this work, N.K.\ has been at the
Max-Planck-Institute for Dynamics and Self-Organization in G\"ottingen.

\pdfbookmark[1]{References}{ref}
\LastPageEnding

\end{document}